\documentclass[12pt,preprint]{aastex}

\shortauthors{Matthews \& Reid}
\shorttitle{\HI\ Observations of AGB Stars}
\received{2006 November 1}
\begin{document}

\newcommand{\msun}{\mbox{${\cal M}_\odot$}}
\newcommand{\lsun}{\mbox{${\cal L}_\odot$}}
\newcommand{\kms}{\mbox{km s$^{-1}$}}
\newcommand{\HI}{\mbox{H\,{\sc i}}}
\newcommand{\CI}{\mbox{C\,{\sc i}}}
\newcommand{\KI}{\mbox{K\,{\sc i}}}
\newcommand{\HA}{H$\alpha$}
\newcommand{\mhi}{\mbox{$M_{\rm HI}$}}
\newcommand{\HII}{\mbox{H\,{\sc ii}}}
\newcommand{\am}[2]{$#1'\,\hspace{-1.7mm}.\hspace{.0mm}#2$}
\newcommand{\as}[2]{$#1''\,\hspace{-1.7mm}.\hspace{.0mm}#2$}
\newcommand{\ad}[2]{$#1^{\circ}\,\hspace{-1.7mm}.\hspace{.0mm}#2$}
\newcommand{\lsim}{~\rlap{$<$}{\lower 1.0ex\hbox{$\sim$}}}
\newcommand{\gsim}{~\rlap{$>$}{\lower 1.0ex\hbox{$\sim$}}}
\newcommand{\dark}{$M_{HI}/L_{B}$}
\newcommand{\tm}{$\times$}
\newcommand{\nan}{Nan\c{c}ay}
\newcommand{\ang}{\rm \AA}

   \title{VLA Observations of \HI\ in the Circumstellar Envelopes of
   AGB Stars}

  \author{Lynn D. Matthews\altaffilmark{1}}
           \author{Mark J. Reid{\altaffilmark{1}}}

\altaffiltext{1}{Harvard-Smithsonian Center for Astrophysics, 
60 Garden Street, MS-42, Cambridge, MA 02138 USA}            

\begin{abstract}
We have used the Very Large Array (VLA) to search for neutral atomic hydrogen
(\HI) in the circumstellar envelopes of 
five asymptotic giant branch (AGB) stars.  We have
detected \HI\ 21-cm emission 
coincident in both position and velocity with the S-type
semi-regular variable star RS~Cnc. The emission comprises a compact,
slightly elongated feature centered on the star with
a mean diameter  of $\sim82''$
($1.5\times10^{17}$~cm),  plus an additional filament
extending $\sim6'$ to the northwest. If this filament is
associated with RS~Cnc, it would imply that a portion of its mass-loss
is highly asymmetric. We estimate 
$M_{\rm HI}\approx 1.5\times10^{-3}M_{\odot}$ and a 
mass-loss rate ${\dot M}\approx1.7\times10^{-7}M_{\odot}$ yr$^{-1}$. 
Toward three other stars (IRC+10216, EP~Aqr, R~Cas), we have
detected arcminute-scale
\HI\ emission features at velocities consistent with the circumstellar
envelopes, but spatially offset from the stellar positions. Toward
R~Cas, the emission is weak but peaks 
at the stellar systemic velocity and overlaps with the location of its
circumstellar dust shell and thus is probably related to the star. 
In the case of IRC+10216, we were unable to
confirm the detection of \HI\ in absorption
against the cosmic background previously reported by Le~Bertre \&
G\'erard. However, we detect arcs of emission at projected distances of
$r\sim14'-18'$ ($\sim2\times10^{18}$~cm) to the northwest of the star. The large
separation of the emission from the star is plausible given its advanced
evolutionary status, although it is unclear if the asymmetric
distribution and complex
velocity structure are consistent with a circumstellar origin.
For EP~Aqr, the detected \HI\ emission comprises multiple clumps
redward of the systemic velocity, but
we are unable to determine unambiguously whether the emission
arises from the circumstellar envelope or from interstellar
clouds along the line-of-sight. Regardless of the adopted distance for
the \HI\ clumps, their inferred \HI\ masses are at least an order of
magnitude smaller than their individual gravitational binding
masses. 
We did not detect any \HI\ emission from our fifth target, R~Aqr 
(a symbiotic binary), but measured a 1.4~GHz continuum flux density of
18.8$\pm0.7$~mJy. R~Aqr is a previously known radio
source, and the 1.4~GHz emission 
likely arises primarily from
free-free emission from an ionized circumbinary envelope.

\end{abstract}



\section{Introduction}
For stars of low to intermediate masses ($0.8\lsim M_{*}\lsim 6 M_{\odot}$),
the asymptotic giant branch (AGB) evolutionary stage is
characterized by significant mass 
loss ($\dot{M}\sim10^{-8}$ to $10^{-4}M_{\odot}$~yr$^{-1}$) through
cool, low-velocity ($\sim$10~\kms) winds. Over the course of roughly
$10^{5}$ years,
the material ejected from the atmospheres of such stars 
forms extensive circumstellar envelopes,
up to a parsec or more in diameter (e.g., Habing 1996). Indeed,
AGB stars are one of the primary means by which processed
material is recycled back into the interstellar medium (ISM). 
Thus knowledge of the mass-loss history of
these stars is key not only to understanding the evolution and
ultimate fate of
low-to-intermediate-mass 
stars (including their transition to the planetary nebula stage), 
but also for constraining the chemical evolution of
the ISM and the factors that govern its small-scale structure. 
In binary systems, the 
material shed by AGB stars also may have an
important influence on the evolution of Type Ia supernovae ejecta
(e.g., Wang et al. 2004; Deng et al. 2004).

To date, most studies aimed at probing the material shed during the
AGB stage have focused on 
trace species (e.g., CO; dust grains; H$_{2}$O, SiO, \& OH masers) due 
to their ready detectability at radio and infrared wavelengths
(e.g., Habing 1996 and references therein). However, 
hydrogen should be by far the dominant constituent of the mass expelled
from AGB stars. Of particular interest is the {\it atomic} hydrogen
component, since unlike molecular species,
\HI\ is not destroyed by
the interstellar radiation field. This implies that observations of
circumstellar \HI\ have the potential
to probe significantly larger distances from AGB stars than studies of
other envelope tracers ($\ge10^{16}$~cm; e.g., Villaver et
al. 2002; Le~Bertre \& G\'erard 2004).

Atmospheric models predict that for the coolest giants 
($T_{eff}\lsim$2500~K), hydrogen in the stellar atmosphere 
will be predominantly molecular, while
for $T_{eff}>$2500~K, 
hydrogen will be mainly atomic (Glassgold \& Huggins
1983). As hydrogen is shed via an outflowing wind, a 
number of additional factors may modify
its state, including the formation of H$_{2}$ on
grains and the dissociation of H$_{2}$ through
chromospheric emission, shocks, a hot companion, or the interstellar radiation
field.  While the relative importance of these various effects remains
largely unknown,  models predict that some
fraction of the hydrogen in AGB circumstellar envelopes should be atomic, 
and consequently, should be observable via the \HI\ 21-cm
line (Clegg et al. 1983; Glassgold \& Huggins 1983; Villaver et al. 2002). 
Furthermore, \HI\ has been detected in emission and/or absorption
toward 
several planetary nebulae (Rodr\'\i guez \& Moran 1982; Altschuler et al. 1986; 
Schneider et al. 1987; Taylor \& Pottasch 1987; Taylor et al. 1989,1990; 
Gussie \& Taylor 1995;
Rodr\'\i guez et al. 2000,2002); the detected material was presumably
expelled by the stars during their AGB stage, although the details of this
process are still poorly understood (e.g., Rodr\'\i guez et al. 2002).

Together the above factors make direct measurements of 
\HI\ radiation of
considerable interest for determining the dominant form of hydrogen in
the circumstellar envelopes of evolved stars
and for helping to constrain the
physical processes governing their evolution.
In addition, \HI\ measurements have the potential to
supply independent assessments of mass-loss rates, terminal
velocities of the stellar wind, and the sizes, structures, and total
masses of the circumstellar
envelopes. 

Despite the abundant motivations for measuring the hydrogen
component in the circumstellar envelopes of AGB
stars, such observations are challenging in practice.
Even for the nearest AGB stars, the
\HI\ signal is expected to be quite weak, and most often is coincident both in
position and frequency with 
strong Galactic foreground and background emission along the line-of-sight.
Indeed, initial efforts to detect \HI\
associated with evolved stars had very limited success. Until
recently, the combined
result of these studies (Zuckerman et al. 1980; Knapp \& Bowers
1983; Schneider et al. 1987; 
Bowers \& Knapp 1987,1988; Hawkins \& Proctor 1992) was only 
one 21-cm line detection of a genuine AGB star
($o$~Ceti=Mira; Bowers \& Knapp 1988), 
along with the detection of one red supergiant
(Betelgeuse$=\alpha$~Ori; Bowers \&
Knapp 1987). These results have often been interpreted as implying that the
material lost from AGB stars must be primarily molecular (e.g.,
Zuckerman et al. 1980; Knapp \& Bowers 1983), although
in many cases, the derived \HI\ upper limits are
not sufficiently sensitive
to rule out atomic hydrogen as an important constituent of the circumstellar
envelope.

This situation has changed dramatically in the past five years.
Using the upgraded \nan\ Radio Telescope, T. Le~Bertre, E. G\'erard, and coworkers 
have recently reported detections of roughly two dozen AGB and related 
stars. Le~Bertre \& G\'erard (2001) first
reported a detection of \HI\ absorption toward 
the carbon star IRC~+10216 (but see
\S\ref{irc10216}). This team subsequently reported \HI\ emission detections
of the semi-regular variable stars
RS~Cnc, EP~Aqr, and X~Her 
(G\'erard \& Le~Bertre 2003; Le~Bertre \& G\'erard 2004; Gardan et al. 2006)
and the carbon-rich semi-regular
variable Y~CVn (Le~Bertre \& G\'erard 2004). Most recently, they
published the results of a 
more extensive survey featuring detections of a variety of types
of AGB stars and planetary nebulae (G\'erard \& Le~Bertre 2006).

These latest results are
tantalizing, as they suggest that at least part of the material in the
circumstellar envelopes of many evolved stars is atomic, and that  
it is feasible to use \HI\ 21-cm emission
as a diagnostic probe of
the late stages of stellar evolution. Indeed, 
the large observed \HI\ extents of the detected envelopes (up to
$\sim$2~pc) confirms that \HI\ probes
different regions of the envelope than CO or other
molecular tracers, and thus can 
trace mass-loss over very large time-scales---up to $\sim10^{5}$~yr. 
That the observed line profile shapes are very different from the
rectangular profiles expected for symmetric,  optically-thin, constant
velocity outflows also suggests that the outflows from the stars are
slowing down with time and that multiple (in some cases highly asymmetric)
mass-loss episodes have occurred. However, a 
more complete interpretation of these discoveries 
requires better
spatial information than can be provided by the $\sim4'$(E-W)$\times22'$(N-S)
beam of the Nan\c{c}ay telescope. Moreover, the interpretation of the
\nan\ spectra depends
on the decomposition of the line profile in the presence of strong
foreground and background emission. Since much of this contaminating
emission is spatially extended, it should 
be resolved out with an interferometer, implying that aperture synthesis
measurements can provide valuable complementary constraints on the
\HI\ line
parameters, and on the existence of circumstellar \HI\ components with a
range of spatial scales. 
Motivated by this, we have recently used the Very Large Array 
(VLA)\footnote{The Very Large Array of the National Radio Astronomy 
Observatory is a facility of the
National Science Foundation, operated under cooperative agreement by
Associated Universities, Inc.} 
to undertake a pilot \HI\
imaging study of five nearby AGB stars. 
The goals of our study included expanding the sample of AGB
stars with sensitive \HI\ observations, as well as
better constraining the sizes and spatial
distributions of the \HI\ envelopes of AGB stars previously detected
with the  \nan\ telescope. 

\subsection{Sample Selection}
For our
pilot \HI\ survey of AGB stars with the VLA, we did not attempt
to study an unbiased sample of objects, but instead 
selected targets that we judged to have a good probability of detection.
Our sample included three stars
for which  single-dish \HI\
detections had been reported at the time our study began
(IRC+10216, RS~Cnc, and EP~Aqr; see above), together with
two additional well-known AGB stars:
R~Cas and R~Aqr. These latter two stars were selected as being 
relatively nearby 
and well-placed in the sky for scheduling purposes. In
addition, both have radial velocities offset
from the peak of the
Galactic interstellar
emission along the line-of-sight (Hartmann \& Burton
1997), and
neither has any strong neighboring 20-cm continuum sources (Condon et al. 1998)
whose sidelobes might complicate
the detection of weak, extended stellar \HI\ signals. 
An additional motivation for targeting R~Aqr is that it is part of a
symbiotic binary system, with the orbit of the hot companion lying 
within the cool giant's circumstellar envelope (e.g., Spergel et
al. 1983; Hollis et al. 1986). 
It has been suggested that 
the primary source of the \HI\ emission detected from the
weakly symbiotic AGB star $o$~Ceti could be
H$_{2}$ photodissociated by its hot companion (Bowers \& Knapp
1988; G\'erard \& Le~Bertre
2003), hence it is of interest to explore whether \HI\ is also 
associated with other symbiotics.
Some basic properties
of our sample are summarized in Table~1.

%
\begin{deluxetable}{lccccclccccc}
\tabletypesize{\tiny}
\tablewidth{0pc}
\tablenum{1}
\tablecaption{Properties of the Target Stars from the Literature}
\tablehead{
\colhead{Name} & \colhead{$\alpha$(J2000.0)} &
\colhead{$\delta$(J2000.0)} & \colhead{$l$} & \colhead{$b$} 
 & \colhead{$V_{\rm sys}$} &  \colhead{$d$}  & 
\colhead{$T_{eff}$} & \colhead{Spectral Type} &\colhead{$\dot{M}$} & 
\colhead{$V_{o}$} & \colhead{Known}\\
\colhead{}     &  & &
\colhead{(deg)} & \colhead{(deg)} & 
\colhead{(\kms)} & \colhead{(pc)} &\colhead{(K)}  & \colhead{}
&\colhead{($M_{\odot}$ yr$^{-1}$)} & \colhead{(\kms)} & \colhead{~Binary?} \\
  \colhead{(1)} & \colhead{(2)} & \colhead{(3)} &
\colhead{(4)} & \colhead{(5)} & \colhead{(6)} & \colhead{(7)} 
& \colhead{(8)} & \colhead{(9)} & \colhead{(10)} & \colhead{(11)} &
\colhead{(12)} 
}

\startdata

RS Cnc & 09 10 38.8 & +30 57 47.3 &194.50 & +42.08 &+7.3 & 122 & 3110
&M6IIIase &(2.3,10)$\times10^{-8}$ & 2.6,8.0 & No \\

IRC+10216 & 09 47 57.4 & +13 16 43.7 & 221.45 & +45.06 &$-$25.5 & 135 &2200 &
C9.5 &$7.4\times10^{-6}$ & 14.6 & No\\

EP Aqr & 21 46 31.8 & $-$02 12 45.9 & 54.20 & $-$39.26 &$-33.4$ &135 & 3236 &
M8IIIvar &(0.23,1.7)$\times10^{-8}$ &1.4,10.8 & No \\

R Aqr & 23 43 49.5 & $-$15 17 04.2 & 66.52 & $-$70.33 &$-28$ &197 &
2800 & M7IIIpevar & $6\times10^{-8}$ & ... & Yes\\

R Cas & 23 58 24.9 & +51 23 19.7 &114.56 & $-$10.62 &$+24.9$ &160 & 2500
& M7IIIe & $1.2\times10^{-6}$ & 12.1 &  No\\

\enddata

\tablecomments{Units of right ascension are hours, minutes, and
seconds. Units of declination are degrees, arcminutes, and
arcseconds. Explanation of columns: (1) star name; (2) \& (3) right
ascension and declination (J2000.0); (4) \& (5) Galactic coordinates;
(6) systemic velocity relative to the Local Standard of
Rest (LSR); (7) distance in parsecs; (8) stellar effective temperature; (9)
spectral type; (10) mass-loss rate, in solar masses per year; 
two values are quoted for cases with multi-component line
profiles; (11) outflow velocity derived from CO observations; two
values are quoted for cases with multi-component line profiles; (12)
single/binary status of star. Coordinates
and spectral classifications were taken from
SIMBAD (http://simbad.harvard.edu).
Mass-loss rates and outflow velocities were taken from
Michalitsianos et al. 1980 (R~Aqr) or from Table~3 of Knapp et al. 1998 
(all other stars).  References for the
remaining quantities are provided in \S~\ref{results}. 
}

\end{deluxetable}

%
\begin{deluxetable}{lr}
\tabletypesize{\scriptsize}
\tablewidth{0pc}
\tablenum{2}
\tablecaption{Summary of Observations}
\tablehead{
\colhead{Parameter} & \colhead{Value}}

\startdata

Array configuration & D \\
Baseline range & 0.035-1.03 km \\
Number of antennas & 26 \\
Observation dates & 2004 July 1 \& 28 \\
Correlator mode & 2AC \\
Bandwidth & 0.78 MHz \\
Channel width (after Hanning smoothing) & 6.1 kHz\\
Velocity separation of channels & 1.29 \kms \\
Velocity center of bandpass (LSR) & 0 \kms \\
Usable velocity range & $-66\le V_{\rm LSR}\le+66$~\kms \\
Primary beam (FWHM) & $\sim31'$ \\

\enddata

\end{deluxetable}

%
\begin{deluxetable}{lcccl}
\tabletypesize{\scriptsize}
\tablewidth{0pc}
\tablenum{3}
\tablecaption{Calibration Sources}
\tablehead{
\colhead{Source} & \colhead{$\alpha$(J2000.0)} &
\colhead{$\delta$(J2000.0)} & \colhead{Flux Density (Jy)} & \colhead{Date}
}

\startdata
3C48$^{a}$ & 01 37 41.2994 & +33 09 35.132 & 15.87$^{*}$ & 2004July28\\

3C286$^{b}$ & 13 31 08.2879 & +30 30 32.958 & 14.72$^{*}$ & 2004July1\\

0958+324$^{c}$ & 09 58 20.9496 & +32 24 02.209 & 1.61$\pm$0.01 & 2004July1\\

1008+075$^{d}$ & 10 08 00.0160 & +07 30 16.552 & 6.00$\pm$0.02 & 2004July1\\

2136+006$^{e}$ & 21 36 38.5862 & +00 41 54.213 & 3.57$\pm$0.01 & 2004July28\\

2321$-$163$^{f}$ & 23 21 01.9589 & $-$16 23 05.187 & 2.19$\pm$0.01 & 2004July28\\

2355+498$^{g}$ & 23 55 09.4581 & +49 50 08.340 & 1.90$\pm$0.01 & 2004July28\\

\enddata

\tablecomments{Units of right ascension are hours, minutes, and
seconds, and units of declination are degrees, arcminutes, and
arcseconds. }
\tablenotetext{*}{Adopted flux density at 1420.5~MHz, 
computed according to the VLA
Calibration Manual (Perley \& Taylor 2003).}
\tablenotetext{a}{Primary flux calibrator for RS~Cnc and IRC+10216.}
\tablenotetext{b}{Primary flux calibrator for EP~Aqr, R~Aqr, and R~Cas.}
\tablenotetext{c}{Secondary gain calibrator for RS~Cnc.}
\tablenotetext{d}{Secondary gain calibrator for IRC+10216.}
\tablenotetext{e}{Secondary gain calibrator for EP~Aqr.}
\tablenotetext{f}{Secondary gain calibrator for R~Aqr.}
\tablenotetext{g}{Secondary gain calibrator for R~Cas.}

\end{deluxetable}


%
\begin{deluxetable}{lccccccc}
\tabletypesize{\scriptsize}
\tablewidth{0pc}
\tablenum{4}
\tablecaption{Deconvolved Image Characteristics}
\tablehead{
\colhead{Source} & \colhead{{$\cal R$}} & \colhead{Taper} & 
\colhead{$\theta_{\rm FWHM}$} & \colhead{PA} &
\colhead{rms} & \colhead{Cont. chan.} & \colhead{Clean Boxes?}\\
      &   & \colhead{(k$\lambda$,k$\lambda$)}
& \colhead{($'\times '$)} & \colhead{(degrees)} & \colhead{(mJy
beam$^{-1}$)} & \colhead{}\\
\colhead{(1)} & \colhead{(2)} & \colhead{(3)} & 
\colhead{(4)} & \colhead{(5)} &
\colhead{(6)} & \colhead{(6)} & \colhead{(7)} }

\startdata

RS Cnc & +5 & ... & \as{54}{5}$\times$\as{53}{2} & 39.0 & 1.7 &
20-40,90-110 & Yes \\

RS~Cnc & +5 & 2,2 & \as{102}{1}$\times$\as{89}{7} & 51.4 & 2.0 &
20-40,90-110 & Yes\\

IRC+10216 & +5 & 2,2 &\as{101}{1}$\times$\as{94}{1} & 60.6 & 1.5 &
15-40,98-119 & No\\

EP~Aqr & +5 & 2,2 & \as{105}{5}$\times$\as{92}{1} & 29.3 & 1.4 &
33-39,94-110 & No\\

R~Aqr & +5 & 2,2 & \as{113}{9}$\times$\as{93}{0} & 13.7 & 1.9 &
10-53,73-110 & Yes \\

R~Aqr (continuum) & +1 & ... & \as{73}{7}$\times$\as{46}{8}& -0.04
&0.38 & ...& Yes \\

R~Cas & +5 & 2,2 & \as{101}{3}$\times$\as{91}{9} & 39.9 & 2.0 & 15-40
& Yes\\

\enddata

\tablecomments{Explanation of columns: (1) target name; (2) robust
parameter used in image deconvolution; $\cal R$=+5 is equivalent to
natural weighting; (3) Gaussian taper applied in $u$ and
$v$ directions, expressed as
distance to 30\% point of Gaussian in units of kilolambda; 
(4) FWHM dimensions of
synthesized beam; (5) position angle of synthesized beam (measured 
east from north); (6) rms
noise per channel (1$\sigma$); (7) channels used for continuum
subtraction; (8) indication of whether or not clean boxes were used during
image deconvolution.}

\end{deluxetable}


\section{Observations\protect\label{observations}}
Our observations were carried out using the 
VLA on 2004 July 1 \& 28. The array was used in
its most compact (D) configuration  (0.035-1.0~km baselines) 
in order to yield maximum sensitivity to 
emission on scales of up to 15$'$. In total, five stars, five phase
calibrators, and two primary flux calibrators were observed.
The July~1 observations (totaling 4 hours)
were obtained
during the day, while the July~28 observations (totaling 5 hours) 
were obtained after
sunset. Total on-source integration times for each target star  
ranged from 75-97 minutes.  Some further details of the observations
are summarized in Table~2.

The autocorrelator was configured in 2AC mode, with
a 0.78~MHz bandpass. After on-line Hanning smoothing, this yielded 
127 channels with  6.1~kHz ($\sim$1.3~\kms) separation in each of 
two independent polarizations
(right and left circular). 

For the observations of all target stars
and phase calibrators, the bandpass was centered at  zero  velocity
relative to the local standard of rest (LSR). 
The primary flux calibrators (3C48=0137+331 and 3C286=1331+305) each were
observed twice; first with
the bandpasses centered at $V_{\rm LSR}=+160~\kms$, and then at 
$V_{\rm LSR}=-160~\kms$,
in order to avoid contamination of the bandpass from
Galactic line emission near $V_{\rm
LSR}\approx0$. The two spectra were then averaged for
calibration purposes.

The Galactic line emission along the lines-of-sight to our phase
calibrators and target stars is sufficiently ubiquitous
that is most likely filled the beams of the VLA antennas and caused some
increase in the overall system temperature, T$_{sys}$. Because our
flux calibrators were observed at velocities free from Galactic
emission, the net result will be a systematic underestimate of the
flux densities of our phase calibrators and program stars. To estimate the severity
of this effect, we examined the \HI\ survey spectra of Hartmann \& Burton
(1997) toward the direction of each of our targets. Since the
Dwingeloo telescope used by Hartmann \& Burton is the same diameter as
the VLA antennas (25m), the mean brightness temperature of the \HI\
emission in these
spectra over the velocity range of our VLA spectral band 
can be directly compared with the
nominal $T_{sys}$ values for the VLA antennas over the same band. 
We estimate the most significant increase in $T_{sys}$ toward the
direction of
R~Cas, where our flux scale may be systematically low by
$\sim$10\%. For the other four program stars the effect is likely
$\lsim$5\%. Because the effects are modest compared with various 
systematic uncertainties, 
we have not attempted to correct the flux densities
quoted in this paper for this effect. 

\section{Data Reduction\protect\label{reduction}}
Our data were reduced using the Astronomical Image Processing System
(AIPS) software. First a ``pseudocontinuum'' data set was produced by
vector averaging the inner three-quarters of the spectral
bandpass. 
The pseudocontinuum data were used to identify and excise
interference, malfunctioning antennas,
and other bad data, and to calibrate the antenna gains. 

Absolute flux levels were established using
observations of standard VLA flux calibrators (3C48 and 3C286),
and antenna phases were calibrated by using observations of bright
point sources interspersed with the observations of each star (Table~2).
The bandpasses of the spectral line data were calibrated using the 
primary flux calibrators.

The  data for two of our targets (RS~Cnc and IRC+10216) 
were obtained during the daytime and 
exhibited some contamination at short spacings, most likely 
from solar interference (the stars were 33$^{\circ}$ and 44$^{\circ}$
from the Sun, respectively).
This short-spacing contamination 
resulted in a low-frequency ripple pattern 
(spatial period $\sim10'$) in all channels of our deconvolved images. 
Since this 
pattern was spatially fixed with a  roughly constant amplitude
from channel to channel ($\sim$2~mJy~beam$^{-1}$), we were able to
effectively remove it  in
the visibility plane, together with the continuum in the fields,
using the AIPS task UVLIN  (Cornwell et al. 1992; see also below).

\subsection{Imaging of the Spectral Line Data}
The spectral-line 
data for each of the target stars were imaged using
the standard CLEAN deconvolution algorithm within AIPS. In all cases we used a
cell size of 10$''$ and imaged a $\sim43'$ region centered on the star. 
To maximize sensitivity to
diffuse, extended emission, natural weighting (robust parameter $\cal R$=+5)
was used in all cases to make an initial image of the line data, 
yielding synthesized beams of $\sim50''$-$60''$. For each target, 
deconvolved images were made first from visibility data with the 
continuum retained in order to identify channels free of line
emission. Subsequently,
UVLIN was used to fit a zeroth
or first-order
polynomial to the real and imaginary components of each
visibility in the line-free 
channels of the $u$-$v$ data 
(see Table~4), and to subtract the continuum before a second,
continuum-subtracted data cube was produced. Use of this method of
continuum subtraction results in a biased noise distribution in the
deconvolved data cube, in the
sense that the channels used to determine the continuum level have
lower RMS noise (in our case
by $\sim$10\%) compared with those excluded from the fit.  We 
therefore made additional test cubes for each
star where different ranges of channels were selected for continuum
fitting in order to ensure that this channel-dependent 
noise did not affect our ability to identify emission features.

To further improve our dynamic range and sensitivity to
extended emission, additional image cubes were computed
using Gaussian tapering, yielding a synthesized beam
with FWHM $\sim100''$ (see Table~4).  Initially, all images were
deconvolved without using any clean boxes (but see below). 
Even with the use  of tapering, 
the effect of missing short spacings is evident in
some of our images.  When relevant, we comment further in
\S~\ref{results} on the implications of this
for our analysis of the large-scale emission from 
individual targets. 

The limited $u$-$v$
coverage  of our observations also led to a second type of artifact in
our data. In all of our initial data cubes (both tapered and untapered),
we found that most channels exhibited a pattern of emission
in the form of broad (several arcminute scale) diagonal stripes and/or
``checkerboard'' patterns. These patterns typically had mean amplitudes of 
$\sim$1-3~mJy beam$^{-1}$ in the tapered image 
cubes, although both the structure and the intensity varied from
channel to channel within each image cube, and roughly 10\% of the 
channels interspersed throughout each cube were completely free of this
effect. The affected channels were not limited to frequencies
with the brightest Galactic interstellar emission, 
although the effect was noticeably stronger in these channels. 
Except for the few channels with the strongest Galactic signals, 
we found this structured noise
could be eliminated if data from the shorter spacings (below
$\sim$0.3k$\lambda$) were excluded during imaging. Detection of this
unwanted emission only on short spacings implies its
source  is likely to be sidelobe contamination of the
primary beam by  large-scale Galactic
emission distributed across the sky that is poorly sampled by the array. 

In order to ensure that the above noise pattern did not produce
spurious signals that might be mistaken for circumstellar emission, 
we made at least two additional image cubes
for each star: one excluding spacings below 0.3k$\lambda$, and another
using data from all spacings, but with clean boxes placed around
any potential circumstellar emission features. We found that
the use of clean boxes also significantly reduced 
the amplitude of the unwanted large-scale 
background pattern. We regard candidate circumstellar \HI\ emission 
features as potentially real only if they were found
to be statistically significant in {\it both} of these additional data
cubes. However, we performed most of our subsequent analysis on data cubes
incorporating the full range of array spacings. 
Table~4 summarizes the properties of the final images used for the
analysis described in this paper.

\subsection{Imaging of the Continuum}
Continuum images of each of our stellar fields were obtained by
first computing a vector average of the visibilities from all 
channels that
were judged to be free of line emission 
(see Table~4). This averaged $u$-$v$ data
set was then imaged
using the AIPS taks IMAGR with a ROBUST weighting parameter of $\cal R$=+1. 
In the case of the RS~Cnc and IRC+10216 fields, additional images with
$u$-$v$ restrictions were imposed in order to suppress the
short-spacing interference described above. Only one of
our target stars, the symbiotic binary R~Aqr, was detected in the
1.4~GHz (21-cm) continuum; those
data are discussed further in \S\ref{raqr}.  None of the other four
stars in our sample have been detected previously at 1.4~GHz, 
and none show detetable 1.4~GHz counterparts on 
the NRAO VLA Sky Survey (NVSS; Condon et al. 1998). This is consistent
with the radio photosphere 
model of Reid \& Menten (1997), which predicts that the 1.4~GHz
emission from
the isolated AGB stars in our sample would be roughly 50-80 times 
weaker than our detection limit (see also Knapp et al. 1994).

\section{Analysis and Results\protect\label{results}}
Below we present the results of our VLA \HI\ survey of five nearby 
AGB stars. In four cases we have detected \HI\ emission
coincident in velocity with the circumstellar envelopes of the target
stars. In each of these cases we discuss the likelihood that the detected
emission is physically associated with the circumstellar
envelope based on the emission morphology, spatial distribution,
velocity structure, and other factors. 
For one of our targets (R~Aqr) we report a detection of the
previously identified 
1.4~GHz continuum emission. Various parameters (or upper limits) 
derived for each star are summarized in Table~5.

\subsection{RS Cnc\protect\label{rscnc}}
\subsubsection{Background}
RS~Cnc is a semi-regular variable star (class SRc?) 
of spectral type MIII8ase and an effective temperature 
$T_{eff}=3110\pm117$~K (Perrin et al. 1998).
RS~Cnc has a chemical type S, and it is believed to
have undergone at least one thermal pulse (G\'erard \& Le~Bertre
2003).
Based on {\it IRAS} 60$\mu$m observations (Young et al. 1993a,b), 
RS~Cnc is known to have an extended circumstellar
envelope in the form of a detached dust shell with
inner radius \am{1}{0} and outer radius \am{5}{8}.

As noted above,
G\'erard \& Le~Bertre (2003) have reported a detection of RS~Cnc in the
\HI\ 21-cm line using the \nan\ telescope. 
Their observed \HI\ profile is
comprised of a broad and a narrow component and is similar in shape to
the ``double wind'' CO
profiles  observed by Knapp et al. (1998).  G\'erard \& Le~Bertre
decomposed their \HI\
profile into a Gaussian component with FWHM$\sim$12~\kms\ and a narrow,
rectangular component with FWHM$\sim$4~\kms. These widths are comparable to
the two components of the CO lines and
are thought to result from two or more distinct
mass-loss episodes.  Using their CO data, Knapp et al. estimated 
mass-loss rates for the two wind components 
of 1.0$\times10^{-7}M_{\odot}$ yr$^{-1}$  and
2.3$\times10^{-8}M_{\odot}$ yr$^{-1}$, respectively,
assuming the {\it Hipparcos} distance\footnote{All
physical quantities quoted in this paper have been scaled to the
distances quoted in Table~1.} to the star of 122~pc.

\subsubsection{Results: RS Cnc\protect\label{rscncresults}}
\begin{figure}[t]
\epsscale{0.7}
\plotone{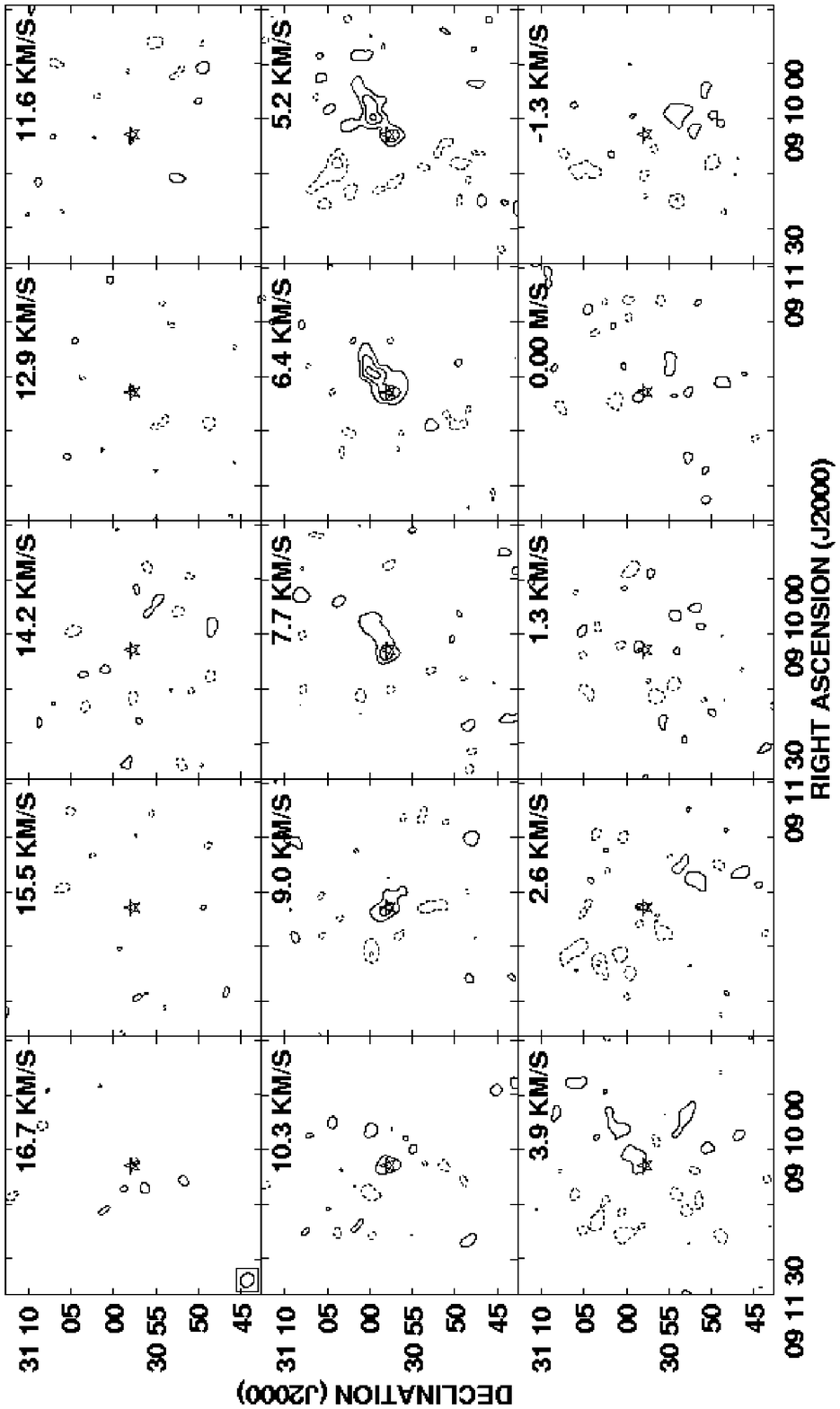}
\figcaption{\HI\ channel maps of the region around RS~Cnc. 
The maps have a spatial resolution of
$\sim102''\times90''$. Contour
levels are (-6,-3,3,6,9)$\times$2~mJy beam$^{-1}$. The lowest
contour levels are $\sim3\sigma$. The systemic
velocity of the star is $V_{\rm sys,LSR}=7.3$~\kms, and a star symbol
marks its optical position. The range of channels shown corresponds to the
velocity range over which CO has been previously detected in the envelope of
RS~Cnc.\protect\label{fig:rscnccmaps}}
\end{figure}

Channel maps from our VLA observations of RS~Cnc are
presented in Figure~\ref{fig:rscnccmaps}.  The channels shown span the velocity
range over which CO emission was detected in the envelope of RS~Cnc 
by Knapp et al. (1998). \HI\ emission is 
clearly detected  ($>5\sigma$) in several
channels bracketing the systemic velocity of the star 
($V_{\rm sys,LSR}=7.3\pm0.3$~\kms\ based on a mean of the CO lines
observed by Knapp et al. 1998). In channels 
spanning the velocity range
$3.9\le V_{\rm LSR}\le 10.3$~\kms, a compact emission feature (hereafter
the ``compact component'') is present,
coincident with the position of RS~Cnc. 
These channels
show an additional, elongated emission component 
(hereafter the ``extended component''), 
extending from the compact feature to roughly $6'$ 
northwest of the star's position, along a position angle of $\sim315^{\circ}$.  
This extended emission is most prevalent between $V_{\rm LSR}$=3.9-7.7~\kms.

Based on  the autocorrelation spectra obtained by each
VLA antenna, we find that
the systemic velocity of RS~Cnc is coincident with modest
Galactic foreground/background 
emission along the line-of-sight. Near the systemic velocity of the
star,  this emission has a brightness temperature
of a few K (see also G\'erard \& Le~Bertre 2003). 
Most of this foreground/background appears to be
resolved out in our aperture synthesis images. 
Therefore the coincidence of the 
compact emission component in both position
and velocity with RS~Cnc strongly suggests a physical association with
the star. The compact
nature of the central component and the lack of any
features of similar strength at other positions in the map or 
in adjacent channels also lend credence
to an association between the \HI\ emission and RS~Cnc. 

To further quantify the significance and uniqueness 
of the detected emission, an automated matched-filter technique was
used to search our data cube for
signals (see Uson \& Matthews 2003). This search was performed on 
the continuum-subtracted, naturally-weighted, untapered data 
cube within a $30'\times30'$ region, over the entire usable
velocity range ($-63$\kms$\le V_{\rm LSR}\le+63$~\kms; i.e., the inner
77\% of the band). 
In essence, signals above a signal-to-noise threshold were identified
by looping through the data,
convolving each spectrum through the designated portion
of the data cube
with Gaussian kernels of width 2-10
channels (2.6-13~\kms). 
The most significant signal was found centered at
$V_{\rm LSR}=7.7$~\kms\ and at the phase center of the data cube,
corresponding to both the position and systemic velocity of
RS~Cnc. The maximum
signal-to-noise of this feature (10$\sigma$) occurred with a convolution width
of six channels ($\sim$8~\kms). Outside of the channels suspected of
containing emission from RS~Cnc (i.e., outside the 
velocity range $3.9\le V_{\rm LSR}\le10.3$~\kms),
the next strongest feature (9$\sigma$) occurred in 
the channel corresponding to the peak of
the Galactic \HI\ signal in the autocorrelation spectra 
(near $V_{\rm LSR}\approx-7.7$~\kms).
In this case the signal is not located at the position of RS~Cnc.
The portion of the data cube searched contained $\sim10^{5}$
independent beams\footnote{This number is not corrected for the
number of trial convolving kernels, since results for the different
kernels are highly correlated.}, hence the probability that the peak signal from
a random \HI\ cloud along the line-of-sight  would
occur at both the position and velocity
of RS~Cnc is roughly 1 in 
$10^{5}$. These results therefore strongly suggest that we
have detected \HI\ emission physically associated with the envelope of
RS~Cnc.

\begin{figure}
\plotone{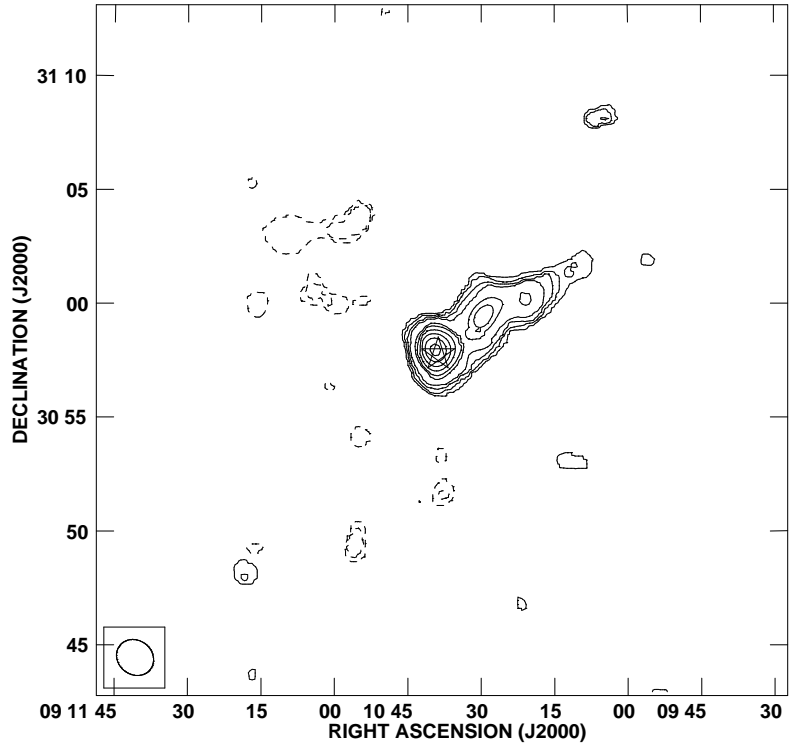}
\figcaption{\HI\ total intensity map of the region around RS~Cnc
derived by summing the emission in channels spanning the velocity
range $3.9\le V_{\rm LSR}\le 10.3$~\kms.
The spatial resolution of the map is $\sim102''\times90''$ and the 
contour levels are
(-16,-8,8,16,24...86)$\times$1.25~Jy beam$^{-1}$ m s$^{-1}$. A star symbol
marks the optical coordinates of RS~Cnc. The size of the region
shown is comparable to that of the FWHM of the primary beam ($\sim30'$). No 
correction for primary beam attenuation
has been applied.\protect\label{fig:rscncmom0}}
\end{figure}

Figure~\ref{fig:rscncmom0} shows an \HI\ total intensity image of the
emission surrounding RS~Cnc, made by summing the emission in channels
spanning the velocity range $3.9\le V_{\rm LSR}\le 10.3$~\kms. 
To increase signal-to-noise, only pixels
whose flux density had an absolute value of
$>2.5\sigma$ after smoothing the data by a factor of three
in velocity were included in the sum. 

To constrain the size of the
compact emission feature coincident with the position of RS~Cnc, we
have used two-dimensional Gaussian fits to the total intensity maps
computed from our naturally-weighted data cubes. In the untapered map, 
the synthesized beam is nearly circular with
FWHM$\sim54''$, while in the tapered map (shown in
Figure~\ref{fig:rscncmom0}) the beam is $\sim102''\times90''$. 
We find that the compact feature is resolved, with a
slight elongation along a roughly north-south direction. As the
position angle of this elongation is not well constrained, we have
fixed it to be 0$^{\circ}$ in our fits. Fitting the tapered image, we
measure deconvolved major and minor axis diameters of $110''\pm11''$
($\sim2.0\times10^{17}$~cm or 0.06~pc) and $54''\pm7''$
($\sim9.8\times10^{16}$~cm or 0.03~pc), respectively. The quoted
uncertainties reflect formal fit errors as well as systematic
uncertainties resulting form varying the size of the box inside which
the fit was performed. Results of the
fits to the untapered image were indistinguishable, but have larger
uncertainties. 
Similar results were also obtained by fitting the
emission in individual channels. The dimensions of the compact \HI\
source are thus comparable to or less than the inner radius of the dust shell
around RS~Cnc found by Young et al. (1993a,b).

\begin{figure}
\plotone{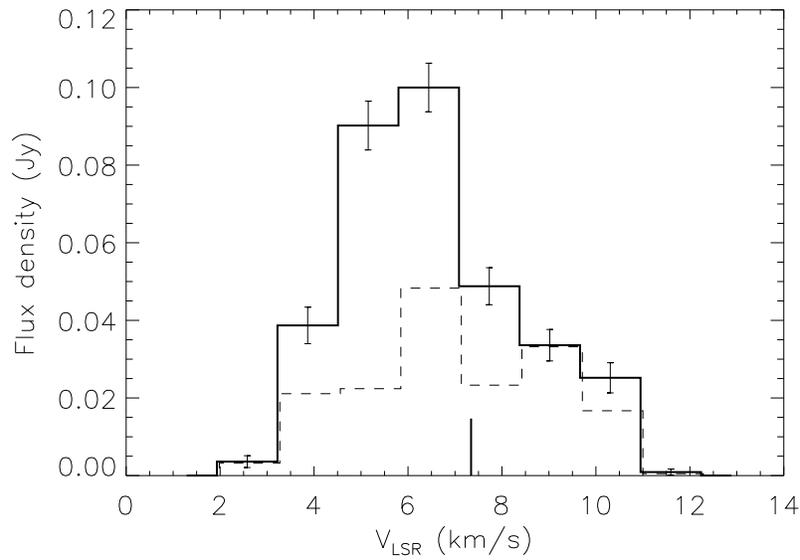}
\figcaption{Global (spatially integrated) \HI\ spectrum of RS~Cnc
derived from VLA observations. The 1$\sigma$ error bars are derived
from the image statistics and do not include calibration
uncertainties. The dashed line shows the profile derived from the
compact emission component only (see Text). 
The vertical bar indicates the stellar systemic
velocity derived CO observations. CO has been detected toward RS~Cnc over 
the velocity interval $0\lsim V_{\rm LSR} \lsim$15~\kms\ (Knapp et al. 1998).
\protect\label{fig:rscncglobal}}
\end{figure}

To estimate the total \HI\ line flux recovered by our VLA
observations, the flux density within each of the channels spanning
the velocity range $2.6\le V_{\rm LSR}\le 11.6$~\kms\
was measured within an irregularly-shaped aperture or ``blotch''. 
For each channel, the aperture was centered near the peak emission
feature and its perimeter was defined 
using the 2$\sigma$ significance contour. All emission within this
aperture was then summed. This analysis was performed on the tapered,
naturally-weighted data cube, after correction for the 
primary beam. The resulting global \HI\ spectrum is shown in
Figure~\ref{fig:rscncglobal}.  The profile shows a slight asymmetry,
with the peak offset by $\sim$1~\kms\ from the stellar systemic
velocity. This offset is significant compared with the formal
uncertainty in $V_{\rm sys}$ as computed from CO observations (see above) and is 
similar to what is seen in the profile of X~Her (another
long-period, semi-regular variable with $T_{eff}>2500$~K) measured by Gardan
et al. (2006), as well as some of the other evolved stars recently detected in
\HI\ by G\'erard \& Le~Bertre (2006). In the case of X~Her,
Gardan et al. suggested that the \HI\ profile morphology must result
from a two-component, asymmetric outflow. 
For RS~Cnc, such an interpretation is
consistent with the compact and extended components seen in
Figure~\ref{fig:rscnccmaps} \& \ref{fig:rscncmom0}. However, an
important difference is that Gardan et al. find the {\it broader}
velocity component of the X~Her profile to be more spatially
extended and linked with a possible outflow along a preferred direction; 
for RS~Cnc we find the FWHM velocity width of the extended emission component to be
narrower than that of the compact component (see below)---although both
components can be traced over comparable velocity ranges. An alternative
explanation for the slight skewing of the line profile may be the effect of
ram pressure as the star moves through the ambient ISM.

From the \HI\ profile in Figure~\ref{fig:rscncglobal}
we measure a peak flux density of $F_{\rm
peak}=0.100\pm0.006$~Jy and an
integrated \HI\ flux of $S_{\rm HI,tot}$=0.44$\pm0.02$~Jy~\kms. 
At the distance of RS~Cnc,
this translates to $M_{HI}=1.5\times10^{-3}M_{\odot}$. The
compact component of the flux distribution centered at the position of
the star comprises approximately half
of the total integrated flux ($S_{\rm HI}\approx0.22$~Jy~\kms). The uncertainty
quoted for the total integrated flux is a formal uncertainty based on
the image statistics,
neglecting calibration uncertainties. After correction for He, the
total envelope mass that we infer
($M\approx2.0\times10^{-3}M_{\odot}$) is comparable to the value derived
from CO and IR data by Young et al. (1993a;
$M=2.2\times10^{-3}M_{\odot}$, assuming a constant mass-loss rate).
However, in
reality, our VLA measurements may underestimate (by a factor
of $\sim$2-3) the total
\HI\ mass of the circumstellar envelope of RS~Cnc if the envelope
contains significant diffuse, extended  
emission. G\'erard \& Le~Bertre (2003) estimated an \HI\ mass of
$\sim1.2\times10^{-3}M_{\odot}$  associated with such an extended 
component, which would nearly double the \HI\ content of the envelope. 

G\'erard
\& Le~Bertre (2003) suggested that the narrow component
of their decomposed \HI\ spectrum of RS~Cnc must arise from an emission region
unresolved by the 4$'$(E-W) \nan\ beam.  
The velocity interval over which we detect emission
near RS~Cnc is similar to the velocity
spread seen in the narrow, rectangular component of G\'erard
\& Le~Bertre's spectrum ($4.5$~\kms$\lsim V_{\rm LSR}\lsim 9.5$~\kms\
based on their Figure~3), and
indeed, roughly one half of the emission detected with the VLA 
lies well within a projected radius around RS~Cnc of $\sim2'$. 
The additional extended component we see in our data to the northwest is
most prevalent over the velocity interval 3.9-7.7~\kms, implying that it is
not the material responsible for the second broad, extended emission component
reported by G\'erard \& Le~Bertre (2003). 
That we do not see any evidence for this broad
component in the VLA data is consistent with the suggestion of G\'erard
\& Le~Bertre  that it
arises from emission extended on scales of at least several arcminutes.

Altough the velocity resolution of our VLA data are too coarse ($\sim$1.3~\kms)
for very precise linewidth determinations, we can 
roughly characterize the \HI\ line shape of RS~Cnc 
as comprising a broad (FWHM$\sim$8~\kms), 
flat component linked with
the compact emission feature, together with a narrower component
(FWHM$\sim$3~\kms) linked to
the extended emission. The broader component has a peak flux density
comparable to the  rectangular component reported by G\'erard
\& Le~Bertre (2003), although the emission we
measure has a slightly broader velocity extent. 
The narrower line component seen in our data
appears to be missing from the composite \nan\ profile, although it
may be blended with the additional broad, spatially extended 
emission component measured by those authors. We also note that a portion of
the emission giving rise to the narrow component of our profile lies 
outside the area subtended by the 
\nan\ beam ($r>2'$). Therefore, part of this emission may not have been
recovered by the position-switched spectra of G\'erard
\& Le~Bertre. To produce their spectra, 
off-beams displaced by $\pm4'$, $\pm6'$, or
$\pm8'$ east and west of the source were
subtracted from the on-source spectra. 
Based on the VLA data, the off-beams obtained $\le6'$ west of the source
would have contained some \HI\ emission, resulting in a net
subtraction from the total flux.

\subsubsection{Discussion: Implications of the Observed \HI\ Emission}
We can estimate a mass-loss rate for RS~Cnc using the \HI\ parameters
derived above. For simplicity, we assume a spherical envelope with a constant
mass-loss rate and constant outflow speed. We emphasize, however, that the actual
situation is likely more complex (see also Gardan et al. 2006). Moreover,
because the relationship between the compact and the extended emission components
is uncertain (see below), we consider only the compact component for this
calculation.  Adopting an \HI\
mass of 0.75$\times10^{-3}M_{\odot}$, 
an outflow speed of 4~\kms\ (HWHM of the \HI\ line from the compact
component) and a diameter of
$1.5\times10^{17}$~cm (the geometric mean of the major and minor
axes), this yields ${\dot M}\approx
1.7\times10^{-7}M_{\odot}$~yr$^{-1}$ after correction for He. 
This rate is intermediate between the values
derived from the two components of the CO profile by 
Knapp et al. (1998). 

The global 
morphology of the extended \HI\ emission we detect in the neighborhood of
RS~Cnc is difficult to explain in terms of a steady,
spherically-symmetric wind. While a roughly spherical 
wind may account for the compact
component of emission centered on the star, the more extended emission
to the northwest 
would seem to require  significant mass-loss
within a rather narrow solid angle. Gardan et al. (2006) have 
argued that outflow along a preferred direction is also necessary to
explain their \HI\ observations of X~Her.
In CO maps, significant deviations from
spherical symmetry are often seen in the circumstellar envelopes of 
AGB stars and have been attributed to 
bipolar outflows (e.g., X~Her, Kahane \& Jura 1996; 
$o$~Ceti; Josselin et al. 2000; IRC+10011, Vinkovi\'c
et al. 2004). The CO emission surrounding 
RS~Cnc is also known to deviate
from spherical symmetry (Neri et al. 1998),
although it is confined to much smaller  radii than the \HI\
($<10''$) and does not appear distinctly bipolar. 
Some type of bipolar phenomenon may account for the slight elongation
of the compact
component seen in our \HI\ maps of RS~Cnc. However, in the case of the
extended component, lack of any 
detectable counterpart to the
southeast argues against such an explantion.

To test whether the extended emission component could instead represent
a bow shock or wake of material produced by ram pressure as the star moves through the
ambient ISM (e.g., Villaver et al. 2003), we
have computed the components of the Galactic peculiar space
motion of RS~Cnc, $(U,V,W)_{\rm pec}$,
following Johnson \& Soderblom (1987). We assumed a heliocentric radial
velocity of +14.4~\kms\ (Wilson 1953), a proper motion in right ascension of $-9.41$
mas yr$^{-1}$, and a proper motion in declination of $-33.05$ mas
yr$^{-1}$ (Perryman et al. 1997). After correction for the solar
motion using the constants of Dehnen \& Binney (1998) we find
$(U,V,W)_{\rm pec}$=($-21,-27,-4$)~\kms. Finally, projecting
back into an Equatorial reference frame, we derive
($V_{r}$, $\alpha$, $\delta$)$_{\rm
pec}$=($18,-17,-24$)~\kms. This predicts a component
of motion toward the southwest (lower right of
Figure~\ref{fig:rscncmom0}), inconsistent with the interpretation of
the extended emission as material swept back by motion through the ISM.
One final possibility we cannot yet discount 
is that the extended \HI\ emission to the
northwest is not associated with RS~Cnc, but 
results from a chance superposition of an unrelated cloud
along the line-of-sight. However, the correspondance between the
radial velocity of this
material and the compact 
\HI\ component centered on the star makes this possibility seem unlikely.

Given the high effective temperature of RS~Cnc (Table~1), the models of Glassgold
\& Huggins (1983) predict that the hydrogen in its upper atmosphere should
be primarily in
atomic form. This suggests that the material we observe was 
shed directly from the stellar atmosphere, rather than 
originating from dissociated H$_{2}$. Estimates of the amount of
dissociated hydrogen expected around the star are consistent with this interpretation.
While the photodissociation rates of H$_{2}$ in circumstellar
envelopes are model-dependent and  still poorly known
(e.g., Morris \& Jura 1983; Glassgold \& Huggins 1983; 
Reid \& Menten 1997), we can obtain a rough
approximation of the maximum number of \HI\ atoms in the RS~Cnc envelope 
resulting from dissociated
H$_{2}$ using the formula suggested by Morris \& Jura (1983). 
Assuming the interstellar UV radiation field is similar to that in the
solar neighborhood (1.9$\times10^{6}$ photons cm$^{-2}$ s$^{-1}$
sr$^{-1}$; Jura 1974), the number of hydrogen atoms in an initially
molecular circumstellar envelope can be approximated as
$18~r^{3}_{\rm max}/V_{\rm out}$,
where $r_{\rm max}$ is the outer radius of the envelope in cm and $V_{\rm out}$ is
its expansion or outflow velocity in \kms.
Taking $r_{\rm
max}=7.5\times10^{16}$~cm (half the mean diameter of the
compact \HI\ component derived
above), $V_{\rm out}\approx$4~\kms\ (HWHM of the \HI\ line), 
this translates to an \HI\ mass for the envelope of
only $1.6\times10^{-6}M_{\odot}$---nearly three orders of magnitude 
smaller than observed. 
While this calculation is crude, the magnitude
of the
discrepancy between the predicted and observed numbers is large and supports a
picture where the bulk of the observed \HI\ emission originated in
the stellar atmosphere of RS~Cnc, not from photodissociated H$_{2}$. 


\subsection{IRC+10216\protect\label{irc10216}}
\subsubsection{Background}
IRC+10216 is the nearest known carbon-rich AGB star
($d\approx$135~pc; Le~Bertre 1997). 
The central star exhibits
a moderately high mass-loss rate 
($\sim 7.4\times10^{-6}M_{\odot}$ yr$^{-1}$;
Knapp et al. 1998) and is believed to be nearing the end of 
its AGB stage, en route to transitioning into 
a planetary nebula (e.g., Gu\'elin et al. 1996; Skinner et
al. 1998).  
The circumstellar envelope of IRC+10216
has been extensively studied at a wide range of wavelengths, and is
known to be structurally complex over a range of scales, 
including an inner, bipolar structure surrounded by numerous arcs and
shells (Gu\'elin et al. 1996; 
Mauron \& Huggins 2000; Fong et al. 2003; Le\~ ao et al. 2006). 
The envelope is also quite extended;  CO emission
has been detected as far as 200$''$ from the star (Huggins et al. 1988),
while the 100$\mu$m emission is present
to $r\sim$\am{9}{5} (Young et al. 1993a,b). 

Zuckerman et al. (1980) and Bowers \& Knapp (1987)
attempted unsuccessfully to detect \HI\ associated with IRC+10216
using the Arecibo telescope and
the VLA, respectively. However, more recently, Le~Bertre \& G\'erard
(2001) reported a detection of this star in \HI\ {\it
absorption}  using the \nan\
Radio Telescope. As we describe below, our
new VLA observations are consistent with the possible existence of atomic hydrogen
surrounding IRC+10216, but the data suggest a possible modified interpretation of
Le~Bertre \& G\'erard's results.

\subsubsection{Results: IRC+10216}
Selected \HI\ channel maps from our VLA observations of IRC+10216 are
shown in Figure~\ref{fig:irccmaps}. These maps reveal 
extended, spatially contiguous patches of 
\HI\ emission within several channels bracketing the 
systemic velocity of IRC+10216 
[$V_{\rm sys,LSR}=-25.5\pm0.3$~\kms\ based on
the CO(2-1) 
line; Knapp et al. 1998]. The velocities of the detected emisson  are consistent with
other tracers of the envelope, including \CI\ (Keene et
al. 1993; van der Veen et al. 1998) and CO (Knapp et al. 1998; 
Fong et al. 2003). However,
no \HI\ emission is
seen directly toward the position of the  star itself; instead the
bulk of the emission is visible at projected
distances of $\sim$\am{14}{3}-\am{18}{0}
to the northwest.  
 
The morphology of the emission detected near IRC+10216 is highlighted further in 
Figure~\ref{fig:ircmom0}, where we present an \HI\ total intensity image
obtained by summing the data from channels spanning the velocity
interval $-31$~\kms$\le V_{\rm LSR}\le -17$~\kms. 
To improve signal-to-noise, only pixels 
having flux densities with absolute values $\ge2.5\sigma$ after
smoothing the data in space and velocity
by a factor of three were
included. It can be seen that the bulk of the detected
\HI\ emission at velocities close to IRC+10216 lies 
along two clumpy, elongated structures,  extending between
position angles $\sim295^{\circ}$-345$^{\circ}$. These position angles
do not have any obvious relation to the geometry of 
features detected at other wavelengths, such as the bipolar structure seen
in the optical (Maurin \& Huggins 2000; Le\~ ao et al. 2006)
or the pattern of arcs seen in CO (Fong et al. 2003).

\subsubsection{Discussion: Have We Detected
\HI\ Emission in the IRC+10216 Envelope?}
Unlike the case of
RS~Cnc, where we detected \HI\ emission coincident both in velocity and in
position with the central star, the data for IRC+10216 are more difficult
to interpret, particularly since we know {\it a priori} very little about
the expected sizes, structures, and morphologies of the \HI\
envelopes around AGB stars of different ages and temperatures. We now
consider evidence for and against a possible association of the
emission we have detected toward IRC+10216 with the envelope of the star.

The Galactic 21-cm foreground/background emission in the direction 
of IRC+10216 is complex, although its peak brightness temperature is modest
(a few K), and its brightness drops to $\sim$1~K near the systemic velocity of 
IRC+10216 
(Hartmann \& Burton 1997; Le~Bertre \& G\'erard 2001). As it is expected that the
bulk of this weak emission should be  
resolved out in the VLA synthesized images, the
likelihood that we have detected emission from an intervening foreground
or background cloud along the line-of-sight to IRC+10216 should be
rather low. However, the 
ISM is known to contain structures on a wide variety of scales,
including arcminute-size structures (e.g., Baker \& Burton 1979;
Greisen \& Liszt 1986; Knapp \& Bowers 1988; Gibson et al. 2000;
Braun \& Kanekar 2005;
see also \S~\ref{epaqr}, \S~\ref{rcas}) and even subarcsecond-scale
features (e.g., Dieter et al. 1979). Moreover, some of these
structures may have velocities
deviating from the underlying Galactic rotation (e.g., Knapp \& Bowers 1988; 
Lockman 2002). Therefore the 
possibility of line-of-sight contamination in our present observations
cannot be immediately excluded.

To aid in assessing the probability that some or all of the \HI\
emission we have detected in our VLA data could be
associated with the circumstellar envelope of IRC+10216,
we have searched our untapered, naturally-weighted IRC+10216 data cube for
signals using the matched
filter technique described in \S~\ref{rscncresults}. After
smoothing the data in frequency with Gaussian kernels of various widths (2-10
channels), 
we find a total of twelve channels containing  ``signals'' with
significance $>7\sigma$ .
Nine of these channels are outside the velocity range where
either CO or
\CI\ emission has been detected in the IRC+10216 envelope; 
these channels (spanning $-2.6\le V_{\rm LSR}\le 1.3$~\kms\ and 
$-10.3\le V_{\rm LSR}\le -5.2$~\kms) also lie
within the velocity range where the Galactic foreground/background
emission along the line-of-sight is strongest based on our
autocorrelation spectra. The emission within these
nine channels is therefore unlikely to be associated with IRC+10216.
In contrast, the other
three channels where $>7\sigma$ features were found
correspond to velocities that bracket (to within uncertainties) 
the systemic velocity of IRC+10216 ($-28.3\le V_{\rm LSR}\le
-25.8$~\kms) and
have underlying Galactic emission that is roughly eight times weaker than in
the nine channels mentioned above. 
This is consistent with a possible
relationship between the emission  and
the envelope of IRC+10216. Nonetheless, we 
note that we targeted IRC+10216 with the VLA specifically because of the
previously reported detection of \HI\ near the velocity of the star
(Le~Bertre \& G\'erard 2001). 
Since a velocity coincidence between the star and \HI\ emission toward 
this direction was therefore expected {\it a priori}, 
these coincidence alone do not strongly discount the 
possibility of line-of-sight contamination. 

\begin{figure}
\epsscale{0.8}
\plotone{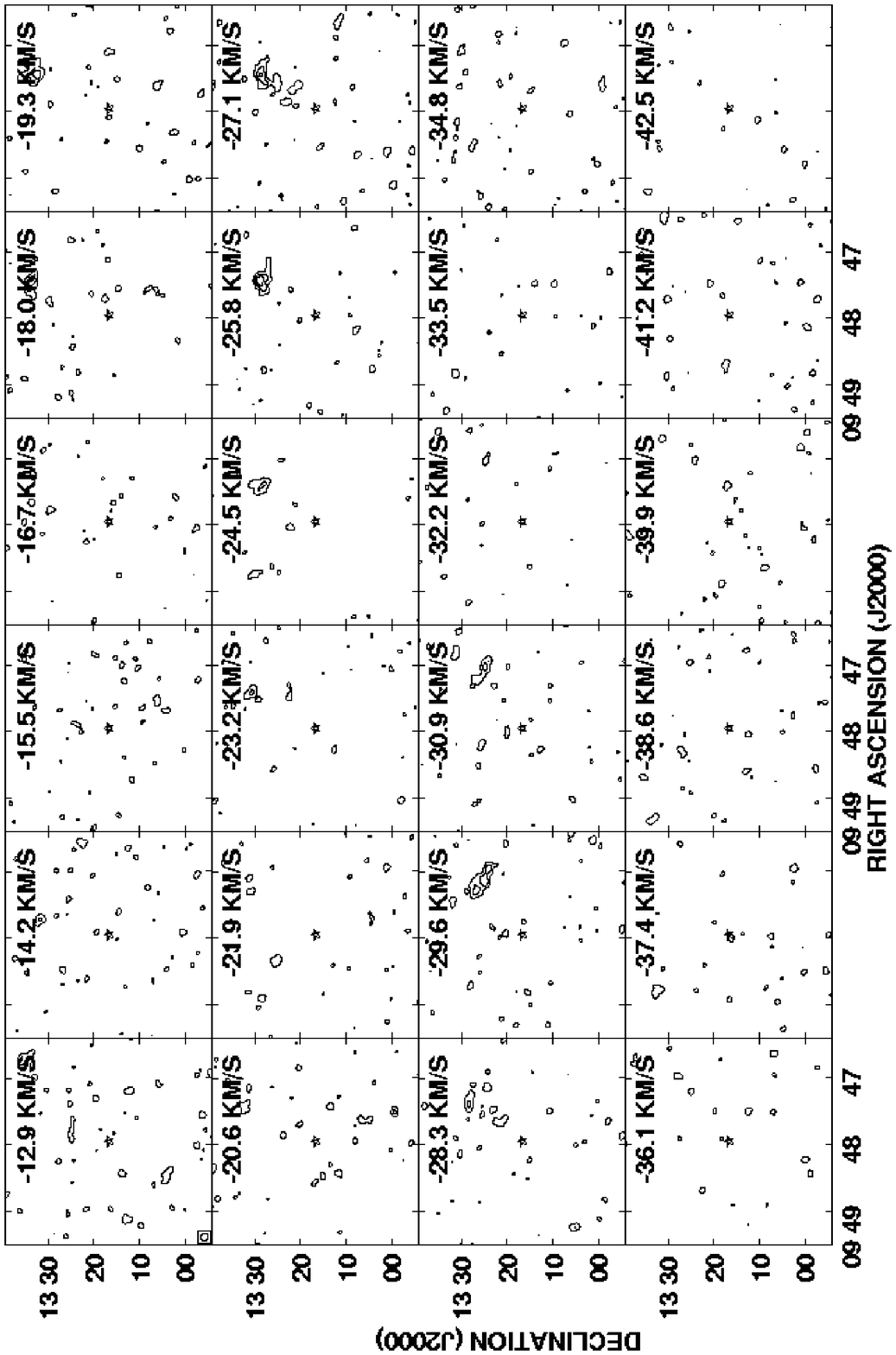}
\figcaption{\HI\ channel maps of the region around 
IRC+10216. The maps have a spatial resolution of
$\sim101''\times94''$. Contour
levels are (-6,-3,3,6,9)$\times$1.5~mJy beam$^{-1}$. The lowest
contour levels are $\sim3\sigma$. The systemic
velocity of the star derived from CO observations 
is $V_{\rm sys,LSR}=-25.5$~\kms, and a star symbol
marks its position. The range of channels shown corresponds to the
velocity range over which CO has been previously detected in the envelope of
IRC+10216.\protect\label{fig:irccmaps}}
\end{figure}

\begin{figure}
\epsscale{0.95}
\plotone{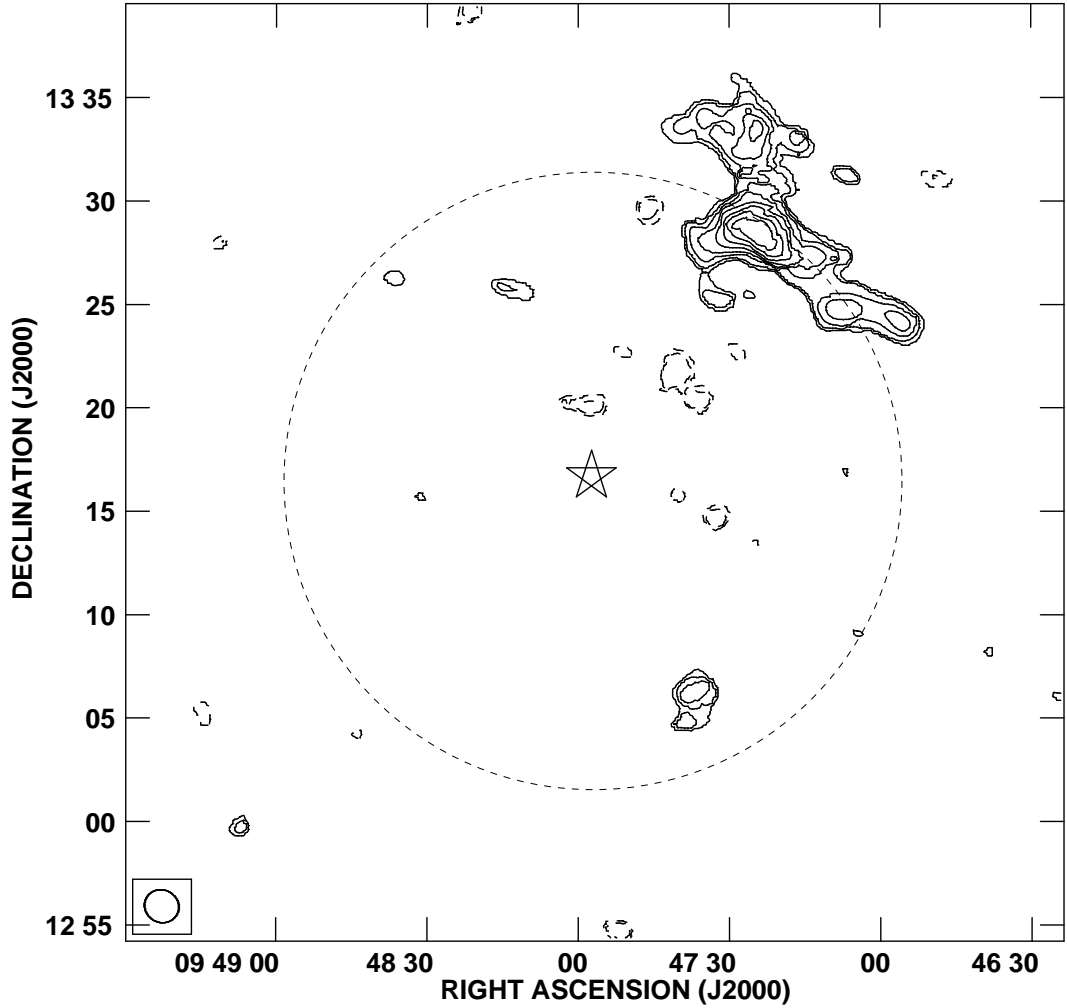}
\figcaption{\HI\ total intensity map of the region around IRC+10216,
derived from data over the velocity range $-31$~\kms$\le V_{\rm
LSR}\le-17$~\kms.
The spatial resolution of the map is $\sim101''\times94''$ and the 
contour levels are
(-12,-6,6,12,18...48)$\times$1.2~Jy beam$^{-1}$ m s$^{-1}$. No correction
for attenuation of 
the primary beam (shown as a dashed circle) has been applied. A star symbol
marks the position of IRC+12016. 
\protect\label{fig:ircmom0}}
\end{figure}

Additional evidence of a possible relationship between the 
\HI\ features in Figure~\ref{fig:irccmaps}
and the envelope of IRC+10216 comes from the geometry of the 
emission and its projected distance from the star.
As noted above, the \HI\ emission detected
in our VLA images lies primarily along two clumpy,
arc-like structures. Figure~\ref{fig:ircmom0} illustrates that the
shape and orientation of these arcs appear roughly
consistent with material lying
along ring- or shell-like structures 
centered on the IRC+10216 central star. 
These \HI\ features also show an  intriguing similarity to 
features
previously seen in the extended {\it molecular} envelope of IRC+10216. 
Using $^{12}$CO(1-0) spectral line observations,
Fong et al. (2003) uncovered evidence for a
series of clumpy, arc-like structures surrounding IRC+10216,
superposed on a smoother, extended 
molecular envelope. They found these molecular arcs to be visible
over the range \am{0}{4}$\le r\le$\am{2}{3} and $-38$~\kms$\le V_{\rm LSR}\le
-14$~\kms. These arcs exhibit a wide range of azimuthal lengths, and
are not symmetrically distributed about the central star.
Analogous shells and arcs have also been found around other 
carbon stars and are believed to be formed through
brief ($\sim$100 yr) but intense periods of mass-loss ($>10^{-5}
M_{\odot}$ yr$^{-1}$; e.g.,
Steffen \& Sch\"onberner 2000; Sch\"oier et al. 2005).

Compared with the CO arcs reported by Fong et al. (2003), the 
\HI\ emission detected with the VLA lies at a significantly larger projected
distance from the star. However, this is consistent with the
predictions of models for the expanding and evolving envelope. Even if hydrogen is
initially shed from the IRC+10216 central star in molecular form, the advanced
evolutionary status of the star implies that some of
this material should have now traveled $\sim10^{18}$~cm or more and 
become largely photodissociated by the interstellar
radiation field (Glassgold \& Huggins
1983; Villaver et
al. 2002). Indeed, model calculations for IRC+10216 predict that the
envelope of IRC+10216 should transition to a primarily atomic
composition near a radius $r\sim10^{18}$cm (Le~Bertre \&
G\'erard 2001)---consistent with the projected radii of the \HI\ arcs seen in
our VLA images (1.7-2.2$\times10^{18}$~cm). 

Using the expansion velocity of the IRC+10216 wind obtained from CO
observations ($V_{\rm out}=14.6$~\kms; Knapp et al. 1998),
we can estimate a kinematic age  of $\sim$45,000 years for 
material located at $2\times10^{18}$~cm from the star.  
This timescale is comparable to the expected separation of thermal pulses
for TP-AGB stars ($10^{4}-10^{5}$ years; e.g., Villaver et al. 2002), 
which have been suggested as a possible origin for 
detached shells around carbon stars. However, Villaver et al. (2002) 
predict that structures
formed via this mechanism would disperse on timescales
$t\lsim20,000$~yr. They propose that longer-lived
shells are created primarily
by shocks between consecutive episodes of mass-loss or by the
continuous accumulation of material in the interaction region between
the circumstellar envelope and the
ISM.  Consistent with the observations of IRC+10216, 
Villaver et al. predict that shells formed via these latter
mechanisms should be observable at
distances of $10^{18}$~cm or more from evolved stars.

At the distance of  IRC+10216,
the total \HI\ mass we infer for the emission detected with the VLA is
consistent both with published models of IRC+10216 and with
observationally-determined shell masses  for other
carbon stars. Using the ``blotch'' method described in
\S~\ref{rscncresults}, we 
derive a integrated \HI\ flux of
$S_{\rm HI}\approx0.55\pm$0.02~Jy~\kms\ for the emission visible in 
Figure~\ref{fig:irccmaps}, corresponding to 
$M_{\rm HI}\approx2.4\times10^{-3}M_{\odot}$. 
This mass estimate has considerable uncertainty, since much of
the  emission lies
outside the FWHP radius of the VLA primary beam, 
and because our data appear to be missing some flux on
short spacings. Nonetheless, our inferred \HI\ mass is comparable to 
that predicted for
IRC+10216 by Glassgold \& Huggins (1983; $M_{\rm
HI}\sim2\times10^{-3}~M_{\odot}$)
and is also
comparable to the gas masses of detached shells
around several carbon stars derived by Sch\"oier et
al. (2005) from
CO observations [(1-8)$\times10^{-3}~M_{\odot}$]. 
It is predicted that these shells will sweep up material from the surrounding
medium as they expand, so the  shell masses may include
material shed via earlier, less energetic winds (Villaver et al. 2002;
Sch\"oier et al. 2005).

While the above findings appear consistent with the possibility of an
association between the \HI\ features detected near IRC+10216 with the
VLA and the circumstellar envelope of the star, it is unclear if other
properties of the emission fit as readily with such a model,
including the confinement of the detected
\HI\ emission to a rather narrow range of position angles 
($\Delta$PA$\sim50^{\circ}$) and 
the velocity structure of the material. While the velocity
spread of the \HI\ emission we have detected with the VLA is consistent
with the velocities of material previously observed in both \CI\ and in
CO, unlike these other tracers, the velocity 
spread of \HI\ emission is
not distributed symmetrically about the systemic velocity of the star,
nor does it extend to the full range of velocities where \CI\ and CO
have been detected. Moreover, the highest velocity \HI\ material is located
at the largest projected distances from the star, unlike what is
expected for a simple expanding shell (see e.g., Fong et al. 2003).

\begin{figure}
\plotone{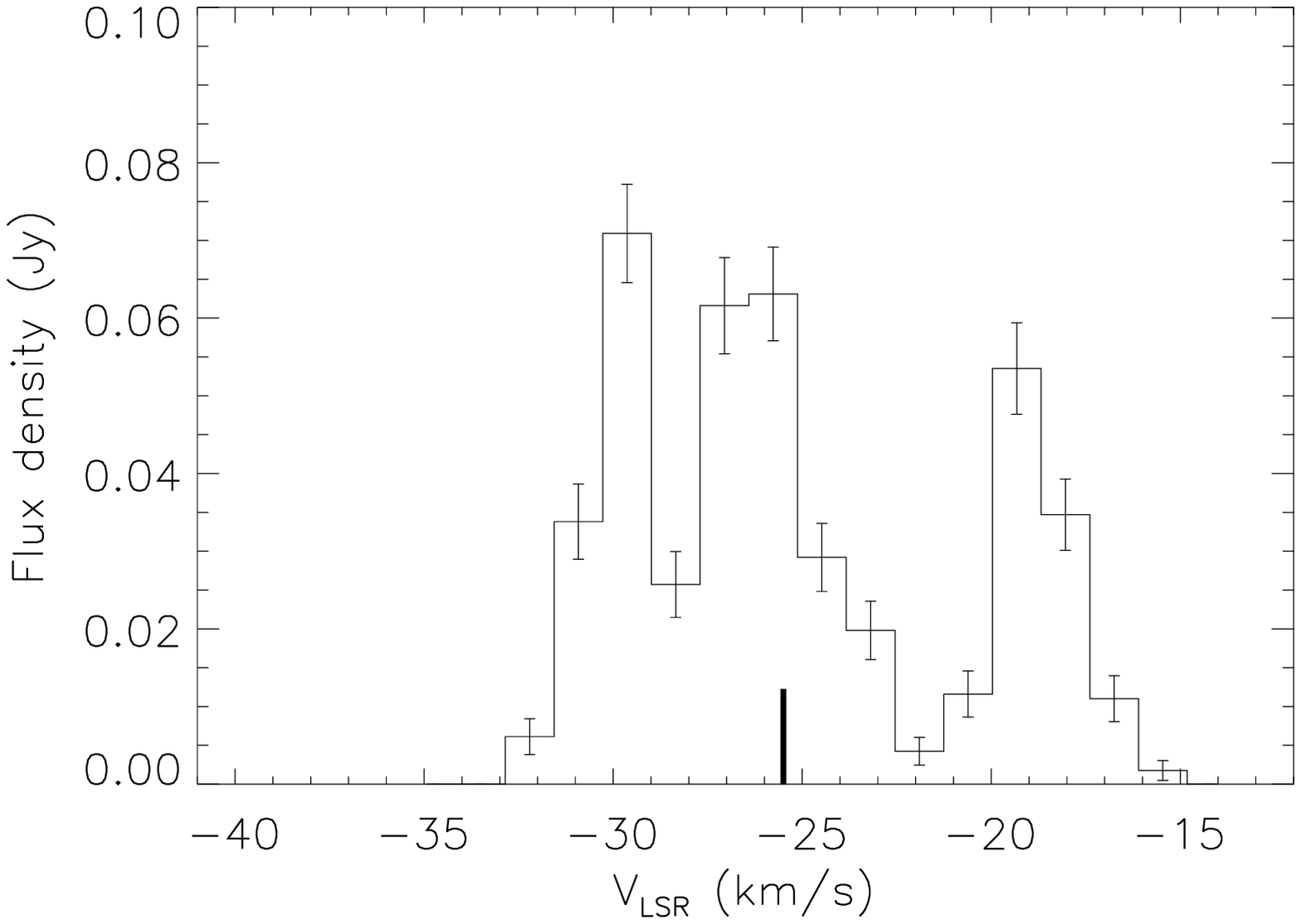}
\figcaption{Global (spatially integrated) 
\HI\ spectrum derived from emission detected toward
IRC+10126. Error bars are as in Figure~\ref{fig:rscncglobal}.
The vertical bar marks the systemic velocity of the star derived from CO 
observations. CO has been detected toward IRC+10216 
over the velocity interval $-52$~\kms$\lsim V_{\rm LSR} \lsim
-$12~\kms\ (Knapp et al. 1998).
\protect\label{fig:ircglobal}}
\end{figure}

The global (spatially integrated) \HI\ profile derived from our VLA data
is shown in Figure~\ref{fig:ircglobal}. 
This profile is noticeably asymmetric
about the stellar systemic velocity. The velocity extent of the
red edge of the profile is comparable to the \CI\ profile measured by
Keene et al. (1993), the CO emission measured by Fong et al. (2003), 
and the \HI\ profile measured by Le~Bertre \&
G\'erard (2001), but compared with these other profiles, the VLA spectrum
shows a dearth of emission on the blue side,  
over the velocity interval $-40$~\kms$\lsim V_{\rm LSR}\lsim -33$~\kms. This
discrepancy is puzzling if a circumstellar origin for the detected
\HI\ emission is assumed.  Possible explanations could be
missing short-spacing flux in the VLA images or blending with
foreground/background emission along the line-of-sight. 
In addition, a significant fraction of the emission detected with the
VLA lies
outside the half-power radius of primary beam, adding further
uncertainty to the global \HI\ profile we have derived. 
An alternative interpretation is that some or all of the emission we
have detected toward IRC+10216 may
be due to superpositions of random 
small-scale \HI\ clouds along the line-of-sight. 
Additional \HI\ mapping of the
area around IRC+10216 
(both with single-dish telescopes and a wide-field VLA mosaic)
is clearly needed to more fully characterize the \HI\ distribution of this
region.

As one final note, we draw attention to some
intriguing similarities exist between the \HI\ emission we have
detected toward IRC+10216 and the \HI\ emission detected in the
Helix planetary nebula.
Rodr\'\i guez et al. (2002) imaged the Helix  in \HI\ using the VLA, 
and discovered 
a large, partial ring-like structure (with $r\sim$0.4~pc)
circumscribing the 
nebula. The \HI\ ring is clumpy and exhibits a
complex velocity structure, somewhat analogous to the emission we see
near IRC+10216, although it is unclear whether 
it could have had a similar origin.
Rodr\'\i guez et al. (2002) suggest the Helix \HI\ ring
is comprised of material released in the form of multiple globules
during the central star's AGB stage, and subsequently photodissociated
by the now hot central star. However, since the central star of 
IRC+10216 is a cool giant,
photodissociation of any ejected globules 
would have to occur primarily via the interstellar radiation field.

\subsubsection{Comparison with Previous Results: 
\HI\ Emission versus \HI\ Absorption?}
One key difference between the new \HI\ observations of IRC+10216 
presented here and those previously reported 
by Le~Bertre \& G\'erard (2001) is that the latter authors reported
seeing only \HI\ in {\it absorption}. They
interpreted this absorption as arising
from supercooled \HI\ ($T_{k}<2.7$~K) in an extended envelope 
seen against the cosmic
background. However,
the analysis of Le~Bertre \& G\'erard was based
on position-switched (on$-$off) 
difference spectra obtained with a single-dish telescope, 
and our VLA data now suggest a possible alternative
interpretation of their results: namely
that the apparent absorption profile 
arose instead from the presence of \HI\ {\it emission}
in the off-source spectra obtained  
to the west of the star. Indeed, Figure~\ref{fig:ircmom0} illustrates that 
any ``off'' spectra taken between roughly $5'$-$15'$ 
to west of IRC+10216 with the highly elongated
\nan\ beam [$\sim21'$(N-S)]
would have sampled the \HI\
emission detected to the northwest of IRC+10216. Meanwhile, little or no emission
would have been detected at the location of the IRC+10216 central
star. Consequently, on$-$off difference spectra derived using spectra
from these two locations would  be expected to exhibit 
negative residuals, thus leading to an apparent absorption feature.
Consistent with this revised interpretation,  Le~Bertre \& G\'erard (2001) reported
seeing stronger absorption signatures in their 
difference spectra using off-beams
displaced by $8'$ and $12'$ from IRC+10216 compared with a
displacement of 4$'$.


\subsection{EP Aqr\protect\label{epaqr}}
\subsubsection{Background}
EP~Aqr is a semi-regular variable (SRb) with a period of $\sim$55 days and a
spectral type of M8III. Dumm \& Schild (1998) derived an
effective temperature for the star of $T_{eff}=$3236~K. 
CO observations have shown that
EP~Aqr has  two-component wind, as evidenced by the presence of both
broad and narrow line components (Knapp et al. 1998; Kerschbaum \& Olofsson
1999; Winters et
al. 2003; Nakashima 2006). Knapp et al. (1998) interpreted the two-component 
CO profiles as indicating
there have been at least two major mass-loss episodes. They
derived mass-loss rates  of 2.3$\times10^{-7}M_{\odot}$ yr$^{-1}$ and 
1.7$\times10^{-8}M_{\odot}$ yr$^{-1}$  from the broad and
narrow line CO components, respectively, assuming a {\it Hipparcos} distance of 135~pc.
The mean central velocity 
of the CO components is $V_{\rm sys,LSR}\approx-33.4\pm0.4$~\kms\ (Knapp et al. 1998), 
and we adopt this 
as the stellar systemic velocity. 

EP~Aqr is known to have a rather extended circumstellar envelope based
on previous infrared observations. Using
on {\it IRAS} 60$\mu$m data, Young et al. (1993a,b) found
that EP~Aqr is surrounded by a detached dust shell of inner radius
\am{1}{5} (0.06~pc) and outer radius \am{5}{9} (0.23~pc). Le~Bertre \&
G\'erard (2004) have also found evidence for an extended envelope
based on their recent \HI\ observations. 

\subsubsection{Results: EP Aqr\protect\label{epaqrresults}}
\begin{figure}
\epsscale{0.9}
\plotone{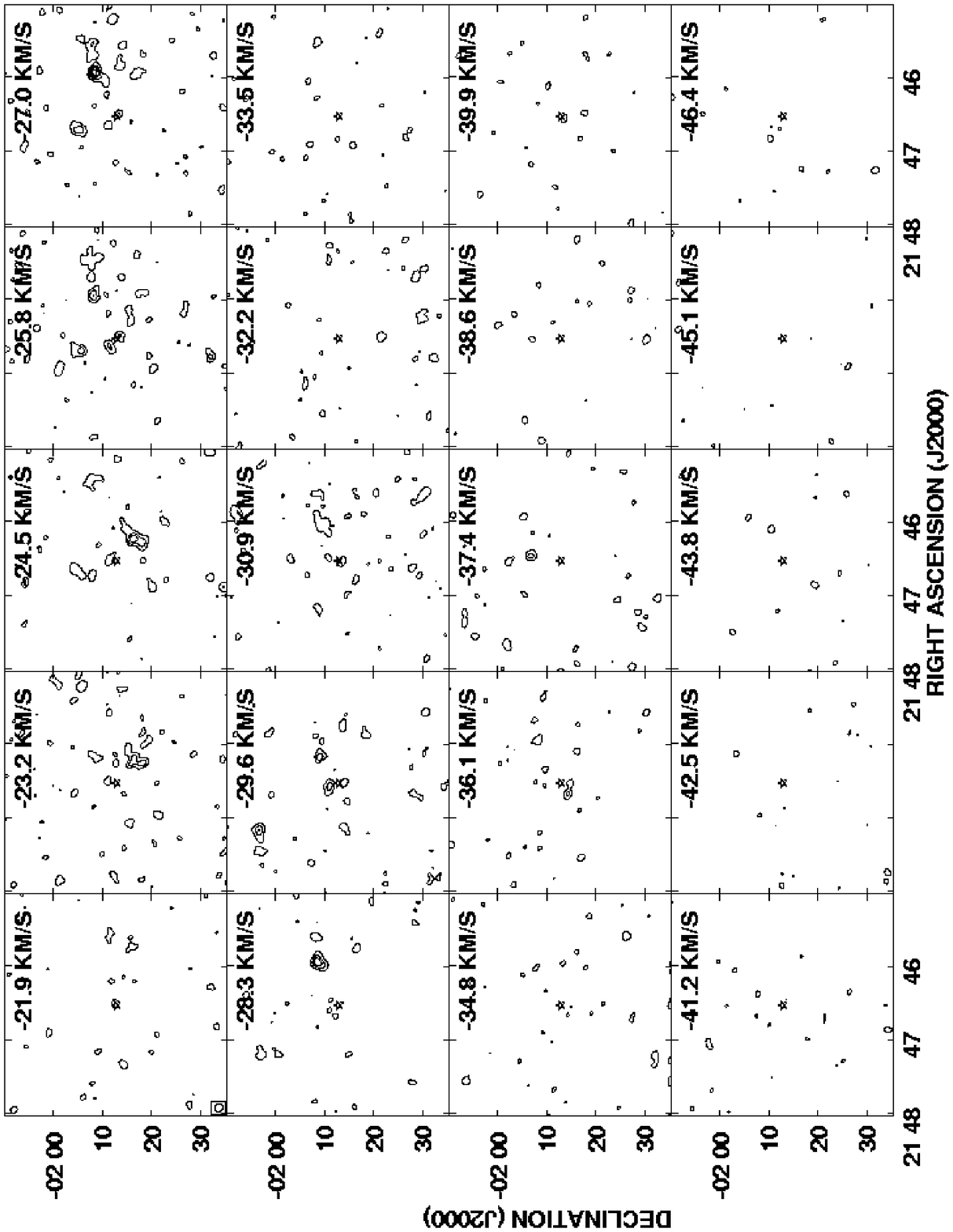}
\figcaption{\HI\ channel maps of the region around 
EP~Aqr. The maps have a spatial resolution of
$\sim105''\times92''$. Contour
levels are (-6,-3,3,6,9,12)$\times$1.5~mJy beam$^{-1}$. The lowest
contour levels are $\sim3\sigma$. The systemic
velocity of the star derived from CO observations 
is $V_{\rm sys,LSR}=-33.4$~\kms, and a star symbol
marks its position. The range of channels shown corresponds to the
velocity range over which CO has been previously detected in the envelope of
EP~Aqr. 
\protect\label{fig:epaqrcmaps}}
\end{figure}

\HI\ channel maps from our VLA observations of EP~Aqr are
presented in Figure~\ref{fig:epaqrcmaps}. The channels shown
correspond to the velocity range over which CO has been detected previously
in the envelope of EP~Aqr ($-46$~\kms$<V_{\rm LSR}<-22$~\kms;
Knapp et al. 1998). No obvious \HI\ emission
features ($\ge5\sigma$) are
seen anywhere in the channel nearest in velocity to 
the systemic velocity of the star, although statistically significant
emission ($>7\sigma$) is visible in several channels redward of the
systemic velocity.

Using the \nan\ telescope,
Le~Bertre \& G\'erard (2004) observed an \HI\ emission profile toward
EP~Aqr that they decomposed into three components, centered at LSR
velocities of $-31.0$, $-26.4$, and $-31.0$~\kms, respectively, and having velocity
widths $\Delta V=$13.0, 2.6 and 1.6~\kms, respectively. Our VLA data
reveal emission that could be related to the latter two components, but
no evidence of the emission responsible for the broadest 
line component. Le~Bertre \& G\'erard (2004) noted that
this broad component appears to arise from a region extended to as
much as 10$'$ from the star, and our VLA data would have
poor sensitivity to emission on these scales. 

\begin{figure}
\plotone{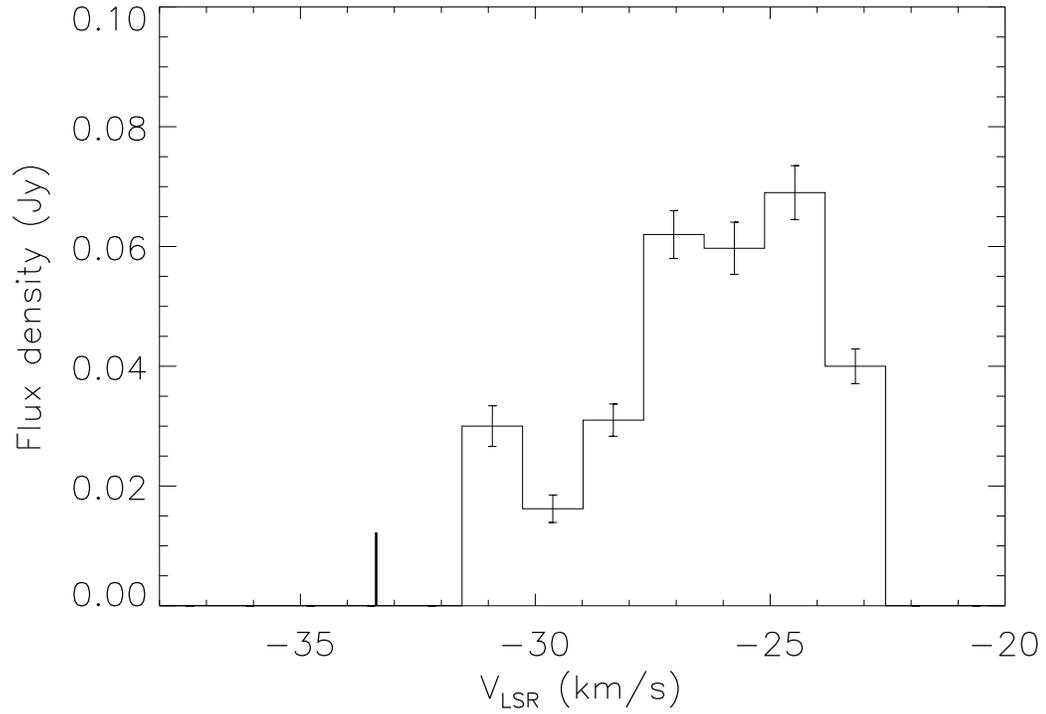}
\figcaption{Global (spatially integrated) \HI\ spectrum  toward
EP~Aqr, derived by summing the emission from clumps
A-F in Table~6. Error bars are as in Figure~\ref{fig:rscncglobal}.
The vertical bar marks 
the systemic velocity of the star derived from CO observations. CO has
been detected toward EP~Aqr over
the velocity interval $-44$~\kms$\lsim V_{\rm LSR} \lsim -$23~\kms\
(Knapp et al. 1998).
\protect\label{fig:epaqrprofile}}
\end{figure}

Le~Bertre \& G\'erard (2004) suggested that the narrowest component seen
in their data (centered at $V_{\rm LSR}=-31$~\kms) 
arises from an emission region
close to the star that is unresolved by their 4$'$(E-W) beam. 
As seen in Figure~\ref{fig:epaqrcmaps}, 
our VLA data do reveal weak ($\sim$4-5$\sigma$)
emission features in channels centered at
$V_{\rm LSR}=-30.9$ \& $-29.9$~\kms, respectively, 
and lying within one-half synthesized
beamwidth from the position of EP~Aqr.
These two channels also show additional, brighter
emission clumps to the northwest, which become more prominent at
higher velocities.  Summing all of this emission yields a 
flux density at these velocities 
comparable to the peak value of 24~mJy reported
by  Le~Bertre \& G\'erard (see Figure~\ref{fig:epaqrprofile}). 
In addition, channels corresponding to velocities in the range
$-30.9$~\kms$\le V_{\rm LSR}\le -23.2$~\kms\ all show multiple
$>5\sigma$ emission clumps that can be traced over two or more
consecutive channels.   
The combined signal from these clumps  likely contributes to 
the intermediate-width emission component that
Le~Bertre \& G\'erard reported centered at 
$V_{\rm LSR}=-26.4$~\kms. 
Le~Bertre \& G\'erard suggested that the material giving
rise to their intermediate-width component must be
concentrated 4$'$-8$'$ northwest of EP~Aqr, and this is roughly consistent
the brightest emission region we detect with the VLA, centered
$\sim$10$'$ northwest of EP~Aqr's position (see Figure~\ref{fig:epaqrmom0}). 

\subsubsection{Discussion: Detection of a Clumpy Circumstellar
Envelope or Background Clouds?\protect\label{epaqrdisc}}
As in the case of IRC+10216 (\S~\ref{irc10216}), we are faced with
some difficulty in assessing whether any or all of the \HI\
emission detected in our VLA observations could be associated with the
circumstellar envelope of our target. 
Compared with the case of IRC+10126, the peak brightness temperature
of the Galactic \HI\ emission along the line-of-sight to EP~Aqr is roughly twice
as strong (Hartmann \& Burton 1997). However, the emission is quite
weak at the velocity of EP~Aqr ($<$1~K; Le~Bertre \& G\'erard 2004), 
and the star is located well
below the Galactic plane ($b=$\ad{-33}{5}), minimizing the
probability for line-of-sight contamination. Nonetheless, as with 
IRC+10216, we specifically targeted EP~Aqr because of a previous
report of \HI\ emission near the systemic velocity of the star, hence
this velocity coincidence alone is not enough to
discount the possibility of foreground/background contamination. 

Within a 15$'$ radius around EP~Aqr,
the bulk of the  emission we detect near the stellar systemic velocity 
is contained within
six fairly compact clumps, $\sim1'$-$5'$ across. 
All of these clumps are detected at
$>5\sigma$ in at least one channel and are traceable over two or more
consecutive channels. We summarize some properties of these clumps in
Table~6. If they lie at the distance of the star, 
the \HI\ masses of these features
are $M_{\rm HI}\sim(0.43$-$14)\times10^{-4}M_{\odot}$. 
The sum of the emission in the clumps with velocities consistent with the
EP~Aqr envelope yields $M_{HI}\approx1.7\times10^{-3}M_{\odot}$.  A global (spatially integrated) \HI\
profile derived by summing this emission is shown in
Figure~\ref{fig:epaqrprofile}. After
correction for He, the inferred mass
is comparable to the envelope mass estimated
from infrared observations ($5\times10^{-3}M_{\odot}$; 
Young et al. 1993b). 

\begin{figure}
\plotone{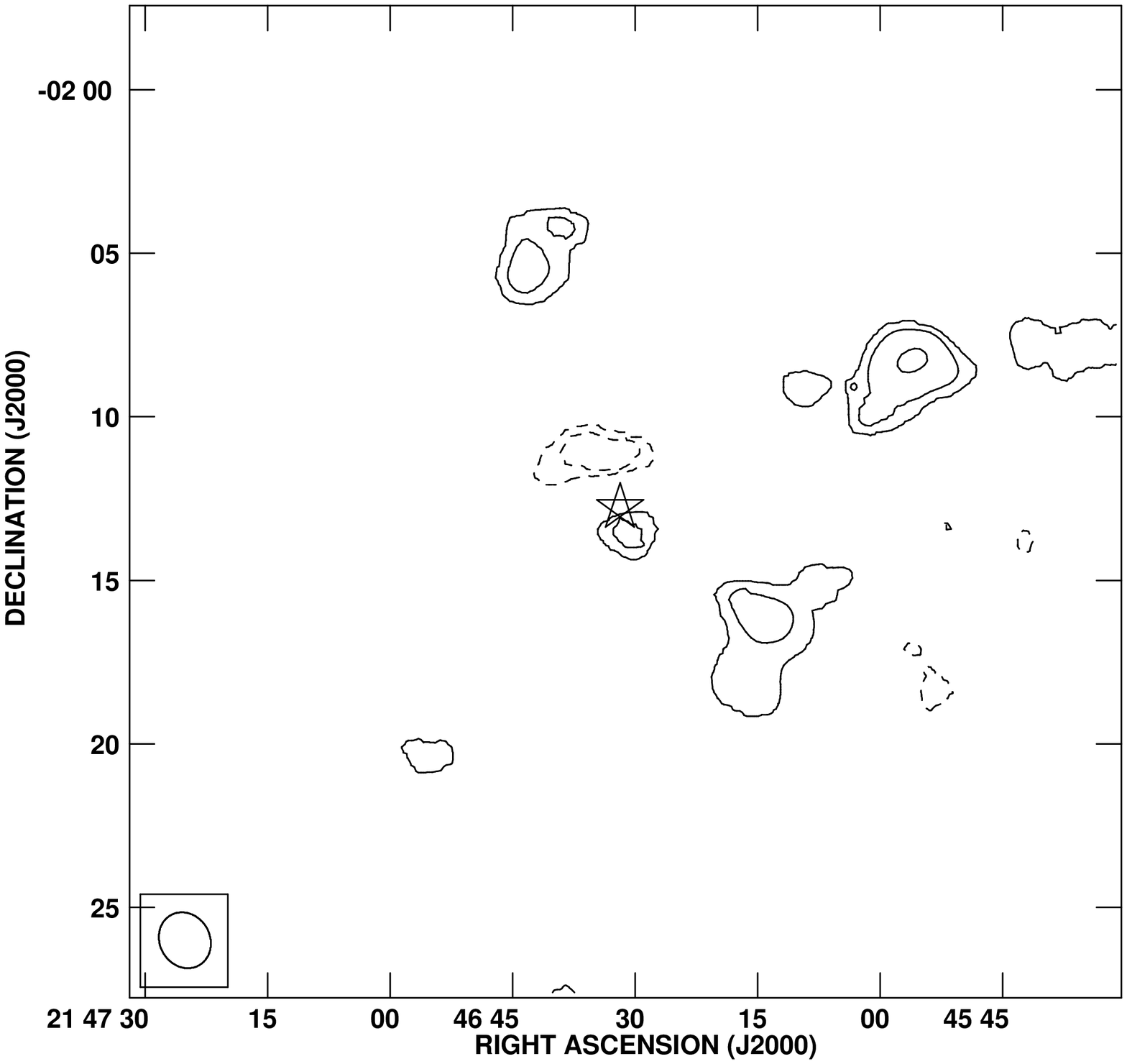}
\figcaption{\HI\ total intensity image of the region around
EP~Aqr, derived from data over the velocity range 
$-32.2$~\kms$\le V_{\rm sys,LSR}\le-23.2$~\kms. 
Contour levels are (-16,-8,8,16,32)$\times$2~Jy
beam$^{-1}$ m s$^{-1}$. A star symbol marks the optical position of
EP~Aqr. The size of the region
shown is comparable to that of the FWHM of the primary beam ($\sim30'$). No 
correction for primary beam attenuation
has been applied. \protect\label{fig:epaqrmom0}}
\end{figure}

Figure~\ref{fig:epaqrmom0} shows an \HI\ total intensity map of a 30$'$
region around EP~Aqr, derived by summing the data over the velocity range
$-32.2$~\kms$\le V_{\rm sys,LSR}\le-23.2$~\kms. Only pixels with
absolute values $\ge2.5\sigma$ after smoothing the data by a factor of
three in velocity were included. 
With the exception of the features denoted C \&
D in Table~6 (which blend together near the center of the image),  
we see no clear evidence that the discrete emission clumps we have detected
are part of any larger, contiguous structures, although three of the
clumps are intersected by a circle with
radius \am{9}{5} centered on the star. 
Moreover, the presence of a strong negative feature near
the center of our map suggests we are missing flux on short-spacings,
so we cannot rule out the existence of a more diffuse, underlying envelope.

The velocities of the emission clumps visible in 
Figure~\ref{fig:epaqrmom0} (see Table~6) are all consistent with 
velocity range over which CO
emission has been detected in the envelope of EP~Aqr, and the range of
\HI\ velocities corresponds with the velocity spread of the ``red
wing'' of its two-component CO profile (Nakashima 2006). However,
whereas the global CO line profiles of EP~Aqr as seen in various
transitions all appear symmetric about $V_{\rm sys}$
(e.g., Knapp et al. 1998; Nakashima 2006), 
all of the \HI\ clumps we have detected are
redward of the systemic velocity. This makes it difficult to
unambiguously establish
a  physical relationship between the detected clumps 
and EP~Aqr's envelope. Moreover, we find an
additional \HI\ clump in our data cube 
with similar properties, but at a significantly  higher velocity
($V_{\rm LSR}=+56.7$~\kms; see Table~6); material at this
velocity is very unlikely to be associated with EP~Aqr's
circumstellar envelope. 

\HI\ clumps with sizes and masses similar to the ones
we find in the direction of EP~Aqr were also
seen in the VLA \HI\ study of the red giant $\alpha$~Ori 
by Bowers \& Knapp (1987; see also Knapp \& Bowers 1988). 
In the case of $\alpha$~Ori,
the velocities of 9 of 13 the detected \HI\ clumps 
are consistent with the
velocity spread of CO emission in the envelope of $\alpha$~Ori 
measured by Knapp
et al. (1998), but the velocities are preferentially concentrated redward of
the stellar systemic velocity. In addition, numerous small CO clouds
were subsequnetly discovered in this direction (Knapp \& Bowers 1988), leading
Bowers \&
Knapp to conclude that the \HI\ clouds they 
detected are unlikely to have a circumstellar origin. 

Assuming the clumps we have detected toward
EP~Aqr are in equilibrium and that they are approximately spherical with uniform
densities, the masses required for them to be gravitationally bound
can be estimated based on the virial theorem:

\begin{equation}
M_{\rm vir}=\frac{5R\Delta V^{2}}{8G{\rm ln}~2}~M_{\odot} 
= 209R\Delta V^{2}~M_{\odot}
\end{equation}

\noindent where $R$ is the radius of the cloud in parsecs, $\Delta V$
is the FWHM velocity width of the cloud in \kms, and the gravitational
constant in these units is $G$=1/232. We have tabulated 
angular size estimates
for the clumps in Table~6. If we assume
the clouds lie at the distance of EP~Aqr and take $\Delta V$ as the
FWHM of the \HI\ line profiles of the clumps, the masses estimated using
Equation~1 range from $<42M_{\odot}$ up to 564$M_{\odot}$.
These values are $\sim$5 orders of magnitude higher than the
inferred \HI\ masses (Table~6). Clearly if the clumps are 
located at the distance of EP~Aqr they cannot be self-gravitating.
Indeed, even if these clumps reside far out in
the Galactic halo ($d\sim$100~kpc), their \HI\ masses would still be
more than an order of magnitude smaller than their virial masses. 
This suggests that
either these clouds are not in equilibrium (and thus are transient
features) or else that the clouds are pressure confined by a
medium that has been resolved out by our interferometric
measurements (see also Knapp \& Bowers 1988). 


\subsection{R Aqr\protect\label{raqr}}
\subsubsection{Background}
After Mira~AB, 
R~Aqr is the nearest known example of a 
symbiotic binary. It is also the closest known example of an
astrophysical jet.  The {\it Hipparcos} distance to this system is
$d$=197~pc, in agreement with the geometric distance of $\sim$200~pc
obtained by Hollis et al. (1997).
R~Aqr is comprised of a primary that is a
Mira-like long-period variable (with a spectral
classification M7IIIpevar and a period of 387 days) and an obscured hot secondary
(most likely a white dwarf with
an accretion disk; e.g., Hollis et al. 2000). 
Although R~Aqr is surrounded by extensive nebulosity (extending to at least
2$'$; e.g., Wallerstein \& Greenstein
1980; Hollis et al. 1985), 
the current mass-loss rate of the star is rather low. Based on
ultraviolet spectroscopy and radio continuum observations, respectively,
both Michalitsianos et al. (1980) and Spergel et al. (1983) 
estimated ${\dot M}\sim6\times10^{-8}M_{\odot}$ yr$^{-1}$.
However, these determinations may underestimate the
actual mass-loss rate may by as much as an order of magnitude,
depending on the fraction of the wind that is ionized 
(e.g., Dougherty et al. 1995).

Given the
effective temperature of the R~Aqr primary ($T_{eff}\approx$2800~K;
Burgarella et al. 1992), the models of
Glassgold \& Huggins (1983) predict that its mass-loss should occur in the
form of atomic rather than molecular 
hydrogen. While a significant fraction of 
the circumbinary envelope of this system is
ionized by the hot companion, the  envelope may contain a
neutral component outside 
its Str\"omgren radius ($r\gsim2.5\times10^{14}$~cm; Kafatos \&
Michalitsianos 1982). At the distance of R~Aqr, 
the lower boundary for this region would lie
well within one VLA synthesized beam.
Models by Spergel et al. (1983)
predict that close to the R~Aqr primary 
($r\lsim4.5\times10^{13}$~cm) 
an additional zone of neutral gas may be present. 
However, if
the ionized nebula is optically thick at 21-cm, any \HI\ gas present
in this zone would be invisible. 
Knapp \& Bowers (1983) previously attempted unsuccessfully to detect R~Aqr
in \HI\ using the VLA, and unlike the other four stars in our present
survey, R~Aqr has never been detected in  CO (Knapp et al. 1989; Young 1995).

\subsubsection{Results: R Aqr Continuum Emission}
Figure~\ref{fig:raqrcont} shows an image of the 1.4~GHz (21-cm) continuum
emission in the R~Aqr field. Approximately 30 continuum sources are
detected within our primary beam, including one corresponding in 
position with R~Aqr. The centroid of this source agrees with the optical
position of the star to within 10$''$ (roughly one-fifth the
width of our synthesized beam) 
implying the continuum emission is associated with R~Aqr.

R~Aqr has long been known to be a radio continuum source, and
detections of the system have been reported previously 
over a wide range of frequencies (1.4-43~GHz; 
e.g., Gregory \& Seaquist 1974; Bowers \& Kundu 1979;
Sopka et al. 1982; Spergel et al. 1983;
Hollis et al. 1985; Dougherty
et al. 1995; Hollis et al. 1997; M\"akinen et al. 2004). 
At high resolution, the continuum emission
breaks up into multiple components, including a compact \HII\ region
centered 
at the position of the AGB star and a jet extended $\sim6''$ to the northeast 
(e.g., Kafatos et al. 1983; Hollis et
al. 1985; Dougherty et al. 1995). The 1.4~GHz continuum from
R~Aqr is much too strong to arise solely from photospheric emission.
As in many other symbiotic systems, the
radio continuum emission from R~Aqr
has been attributed primarily to optically thick free-free emission
from  circumbinary 
material ionized by the hot companion (see Seaquist et al. 1984),
although high resolution images show that there is also a component of
optically thin thermal emission arising from the jet (Hollis et al. 1985).

The jet and \HII\ region components of R~Aqr
seen in the higher resolution 1.4~GHz observations of Hollis et
al. (1985) are unresolved by our current observations, where R~Aqr
appears only as a single point source.
Using an elliptical Gaussian fit, 
we measure a 1.4~GHz continuum flux density for the R~Aqr system of
$F_{\rm 1.4GHz}=18.8\pm$0.7~mJy from our
VLA data. This agrees to within formal uncertainties 
with the 1.4~GHz flux density we derive
from the NVSS survey  (Condon et al. 1998) using the same
method  ($F_{\rm 1.4GHz}=19.1\pm$0.9~mJy), but is significantly higher than
the value previously reported for the \HII\ region+jet by Hollis et al. (1985;
$F_{\rm 1.4GHz}=7.86\pm1.02$ mJy). The Hollis et al. measurements were 
based on higher angular resolution VLA
observations ($\theta_{\rm FWHM}\sim4''$) and therefore may have
resolved out some of the flux.
From our fits we
place an upper limit on the (deconvolved) 
diameter of the R~Aqr radio source $<24''\pm1''$ ($<7.1\times10^{16}$~cm or 
$<$0.02~pc), although the higher resolution observations of Hollis et
al. (1985) have already constrained the source size to be at least several
times small than this. 

\subsubsection{Results: Limits on \HI\ Emission and Absorption in R Aqr}
The stellar systemic velocity of R~Aqr is rather uncertain; the star
has never been detected in CO, and various other
emission and absorption lines yield values that differ by up to
tens of \kms\ and in some cases show variations with time 
(see Wallerstein \& Greenstein
1980). Here we adopt $V_{\rm sys,LSR}\approx-28$~\kms\ based on the \KI\
measurements of Wallerstein \& Greenstein (1980).

Figure~\ref{fig:raqrcmaps} shows the continuum-subtracted
channel images from our spectral line data cube 
over the velocity interval within roughly $\pm10$~\kms\
of the stellar systemic velocity. We see no evidence for 
significant \HI\ emission at or near $V_{\rm sys}$. 
Channels corresponding to $V_{\rm LSR}=-20.6$~\kms\ \&
$V_{\rm LSR}=-23.2$~\kms, respectively, each show a single arcminute-scale emission
feature with significance of $\sim6\sigma$. However, neither feature
coincides with the position of the R~Aqr continuum source 
(as would be expected if \HI\ is
present just outside the Str\"omgren radius; see above) and neither feature 
can be traced beyond a single channel. Neither is therefore a
compelling candidate for emission associated with the circumbinary
envelope.

To estimate an upper limit on the \HI\ content of the 
circumbinary envelope of R~Aqr  
we assume that the neutral portion of the envelope would most likely
be centered at the position of the R~Aqr continuum source and be
unresolved by our beam 
($r<1.7\times10^{17}$~cm). Taking a fiducial velocity extent 
of $\Delta V=10$~\kms,
we then derive a 3$\sigma$ upper limit on the integrated \HI\
flux from R~Aqr as $S_{\rm HI}<3\sigma_{b}\Delta V$ where $\sigma_{b}$ is the
mean rms noise per channel within one synthesized beam
centered on R~Aqr  over the velocity interval
$-33.5\le V_{\rm LSR}\le-23.2$~\kms. 
This yields 
$S_{\rm HI}<0.053$~Jy~\kms, translating to an upper limit on the \HI\ mass of
$M_{\rm HI}<4.9\times10^{-4}M_{\odot}$. Of course this limit does not account
for the possibility of spatially extended \HI\ emission. 

\begin{figure}
\plotone{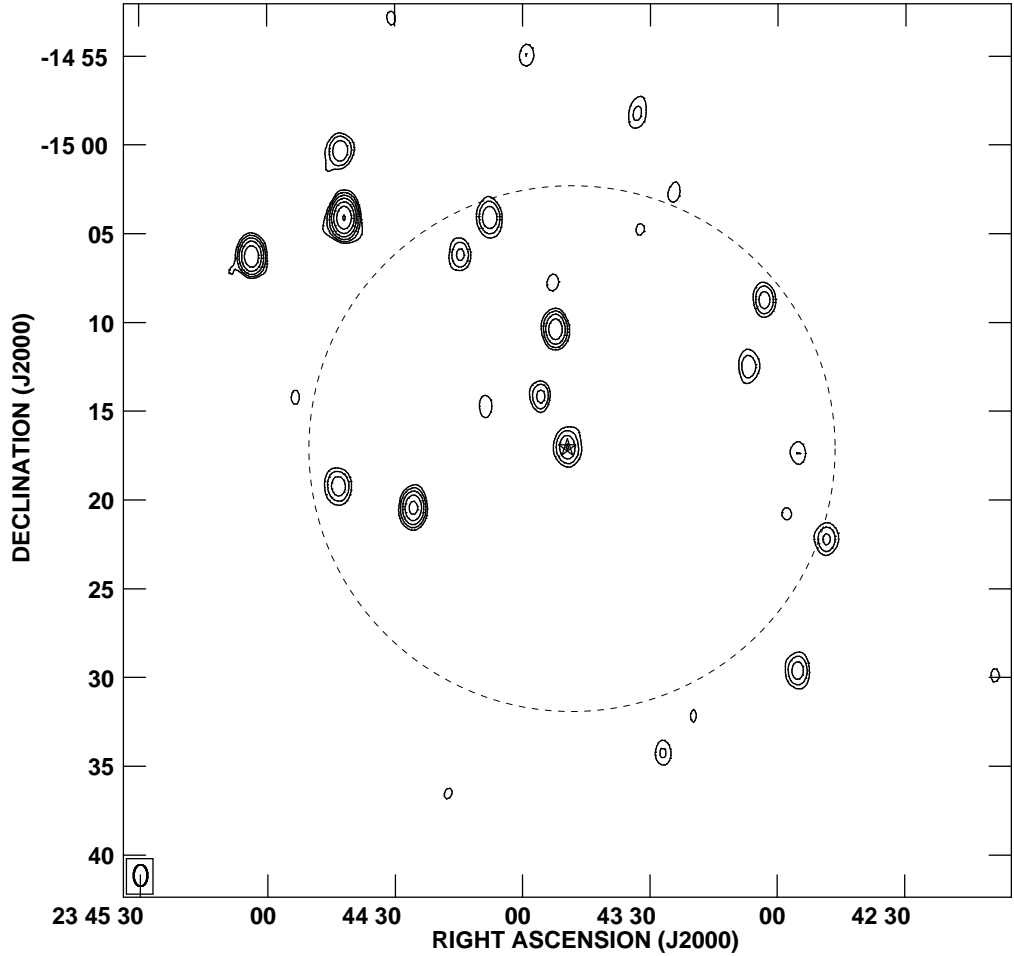}
\figcaption{1.4~GHz Continuum image of the field around R~Aqr. R~Aqr is
detected at the center of the image. Contour levels are
(-5[absent],5,10,20,40,80,160,320)$\times$0.38~mJy beam$^{-1}$.  The
lowest contour levels are 5$\sigma$. The angular
resolution is $\sim74''\times47''$. A star symbol marks the 
optical position of R~Aqr. No
correction  for the primary beam (shown as a dashed circle) 
has been applied.\protect\label{fig:raqrcont}}
\end{figure}

\begin{figure}
\epsscale{0.9}
\plotone{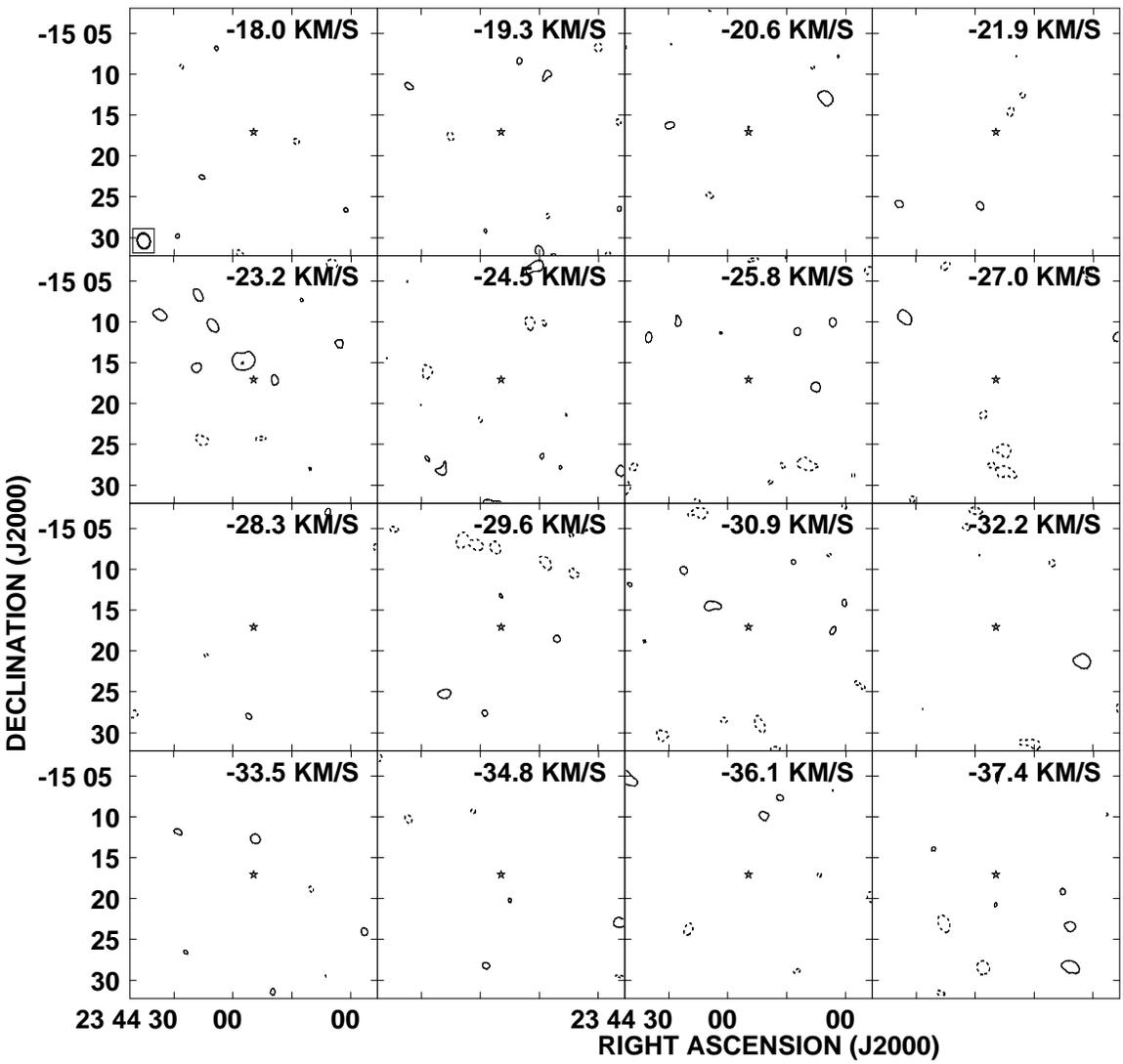}
\figcaption{\HI\ channel maps of the region around 
R~Aqr. The maps have a spatial resolution of
$\sim113''\times93''$. Contour
levels are (-3,3,6)$\times$1.9~mJy beam$^{-1}$. The lowest
contour levels are $\sim3\sigma$. The systemic
velocity of R~Aqr is $V_{\rm sys,LSR}\approx28$~\kms, and a star symbol
marks its position. \protect\label{fig:raqrcmaps}}
\end{figure}

To place additional limits on the possible presence of such an extended
\HI\ envelope, we have also searched for \HI\ in absorption
against the 
background continuum sources seen in Figure~\ref{fig:raqrcont}, 
including the continuum from
R~Aqr itself. However, 
we find no statistically significant \HI\ absorption features
in our bandpass toward any of these sources. This is not surprising, 
as all of the continuum
sources are rather faint ($S_{c}\le0.36$~Jy). We have computed 3$\sigma$
upper limits on the mass of intervening \HI\ along each sightline as:

\begin{equation}
M_{\rm abs,HI}<2.14\times10^{-6}\left(\frac{T_{ex}}{\rm K}\right)
\left(\frac{d}{\rm kpc}\right)^{2}\left(\frac{\theta}{\rm arcsec}\right)^{2}
\left(\frac{\int\tau_{3\sigma} dv}{\rm km~s^{-1}}\right)
\end{equation}

\noindent (Schneider et al. 1987) where
$T_{ex}$ is the excitation temperature of the \HI\ line (taken to be
100~K), $d$ is the distance to the star, $\theta_{c}$ is the angular size
of the continuum source, and $\tau_{3\sigma}$ is the optical
depth of the line at velocity $v$, which can be computed as $\tau_{3\sigma}=-{\rm
ln}\left(1 - \frac{3\sigma}{fS_{\rm c}}\right)$. In the latter
expression, $f$ is the
source covering factor, $\sigma$ is the rms noise, 
and $S_{\rm c}$ is the flux density of the continuum source. 
If we assume that the line has a Gaussian shape, then
$\int\tau_{3\sigma}dv=1.06\tau_{3\sigma}\Delta V$ where $\Delta V$ is
the FWHM linewidth (e.g., Lane 1999). 

Adopting $f=1$ and $\Delta V\approx$10~\kms, toward the R~Aqr
continuum source we find $M_{\rm HI}<0.30M_{\odot}$. The
brightest continuum
source in Figure~\ref{fig:raqrcont} (lying at
$\alpha_{J2000}=23^{h}44^{m}41.9^{s}$, $\delta_{J2000}=-15^{\circ}04^{'}$\as{05}{7}) 
has a flux density of 0.363$\pm0.001$~Jy and yields an upper limit of $M_{\rm
HI}<0.013M_{\odot}$. These absorption limits are thus not sufficiently
strict to rule out the presence of either a compact 
\HI-rich envelope comparable in mass to that seen around RS~Cnc
(\S~\ref{rscncresults}) or a more extended
envelope comparable to those now reported around several other AGB
stars (e.g., G\'erard \& Le~Bertre 2006).


%
\begin{deluxetable}{lcccccc}
\tabletypesize{\scriptsize}
\tablewidth{0pc}
\tablenum{5}
\tablecaption{\HI\ Properties of the Circumstellar Envelopes Derived
from VLA Observations}
\tablehead{ \colhead{Source} & 
\colhead{$M_{\rm HI}$ ($M_{\odot}$)} & 
\colhead{$\theta_{e} (')$} & \colhead{$r_{e}$ ($10^{17}$ cm)} & 
\colhead{$V_{\rm out}$ (\kms)} & \colhead{$V$ (\kms)} &
\colhead{${\dot M} (M_{\odot}$ yr$^{-1}$)} \\
\colhead{(1)} & \colhead{(2)} & \colhead{(3)} & \colhead{(4)} &
\colhead{(5)} & \colhead{(6)} & \colhead{(7)} }

\startdata

RS~Cnc$^{a}$ & 7.5$\times10^{-4}$ & \am{0}{9}$\times$\am{0}{45} 
& 0.5-1.0 & $\sim$4 & 7.7 & 1.7$\times10^{-7}$\\

RS~Cnc$^{b}$ & 7.5$\times10^{-4}$ & 6$'$ & 5.5 & $\sim$1.5 & 7.7 &
... \\

IRC+10216$^{c}$ & 2.4$\times10^{-3}$ & \am{14}{3}-\am{18}{0} & 17-22 &
$\sim$6.5 &$-24.5$ &... \\

EP~Aqr$^{c}$ & 1.7$\times10^{-3}$ & 10$'$ & 12 & ...& $-27.0$ & ... \\

R~Aqr$^{d}$ & $<4.9\times10^{-4}$ & ... & ... & ... & ... & ...\\

R~Cas & 5.3$\times10^{-4}$ & \am{4}{7} & 6.7 & $\sim$2 &24.5 & ... \\

\enddata

\tablenotetext{a}{Parameters for the compact emission component
(see Text).}
\tablenotetext{b}{Parameters for the extended emission component (see Text).}
\tablenotetext{c}{Evidence for an 
association between the observed \HI\ emission and the
circumstellar envelope remains inconclusive.}	
\tablenotetext{d}{Mass limit computed within one synthesized beam
centered on the star.}

\tablecomments{Explanation of columns: (1) star name; (2) \HI\ mass or
3$\sigma$ upper limit; (3) maximum angular extent of
the detected \HI\ emission relative to the position of the star; 
(4) projected linear
extent of the detected \HI\ emission relative to the star; (5) outflow velocity,
based on the HWHM of \HI\
profile; (6) unweighted central velocity (LSR) of the \HI\ emission profile;
(7) mass-loss rate (corrected for the mass of He) estimated from the \HI\ data 
assuming a spherical geometry and a 
constant velocity wind.}

\end{deluxetable}


%
\begin{deluxetable}{ccccllcrcc}
\tabletypesize{\tiny}
\tablewidth{0pc}
\tablenum{6}
\tablecaption{Properties of \HI\ Features Detected Near EP Aqr}
\tablehead{
\colhead{Label}  & \colhead{$\Delta$(Ch.)} 
& \colhead{$V_{0}$}
& \colhead{$F_{\rm peak}$}  &\colhead{$\alpha$(J2000.0)} &
\colhead{$\delta$(J2000.0)} &
 \colhead{$\int S{\rm d}v$} & \colhead{Size} & \colhead{$M_{\rm HI}$} &
\colhead{$M_{\rm vir}$}\\
\colhead{} & \colhead{}  & \colhead{(\kms)} & 
\colhead{(mJy beam$^{-1}$)}  &
\colhead{} & \colhead{} & \colhead{(Jy \kms)} & \colhead{(arcsec)} &
\colhead{($10^{-4}M_{\odot}$)} & \colhead{($M_{\odot}$)} \\
\colhead{(1)} & \colhead{(2)} &\colhead{(3)} &\colhead{(4)}
&\colhead{(5)} 
&\colhead{(6)} &\colhead{(7)} &\colhead{(8)} &\colhead{(9)}
&\colhead{(10)} }

\startdata

A &  84-88 &  $-27.0$ &  20.3 & 21 45 55.8 & $-$02 08 16 &
0.16 & $129''\times66''$ & 6.9 & 100\\

B & 82-84 &  $-24.5$ & 16.8 & 21 46 13.2 & $-$02 18 16 & 0.12
&  $283''\times173''$ & 5.2 & 132\\

C  & 84-85 &  $-24.5$ & 9.7 & 21 46 29.2 & $-$02 13 26 &
0.01 & $<105''\times92''$ & 0.43 & $<$42\\

D  & 87-88 &  $-30.9$ &  8.2 & 21 46 33.2 & $-$02 13 26 & 0.01
& $<105''\times92''$ & 0.43 & $<$42  \\

E  & 83-85 &  $-27.0$ &  12.9 & 21 46 43.2 & $-$02 05 06 &
0.06 & $151''\times79''$ & 2.6 & 134\\

F  & 82-84 &  $-24.5$ &  6.9& 21 46 53.9 & $-$02 20 16 &
0.03 & $<105''\times92''$ & 1.3 & $<$25  \\

G  & 18-23 &  +56.7&  16.6 & 21 45 12.4 & $-$02 18 46 & 0.32
& $228''\times97''$ & 14 & 564\\

\enddata

\tablecomments{\HI\ properties of features were derived from the
naturally-weighted, tapered data cube (see Table~4). Units of
right ascension are hours, minutes, and seconds. Units of
declination are degrees, arcminutes, and
arcseconds. Explanation of columns: (1) clump designation; 
(2) range of channels where clump was detected at $>3\sigma$; 
(3) unweighted mean velocity of clump; 
(4) peak brightness of clump; 
(5) \& (6) right ascension and declination of the  brightness peak; 
(7) velocity integrated flux; (8) approximate angular dimensions of clump;
(9) \HI\ mass of the clump at the distance of EP~Aqr ($d$=135~pc); (10)
gravitational binding mass of the clump at the distance of EP~Aqr.}

\end{deluxetable}

\subsection{R Cas\protect\label{rcas}}
\subsubsection{Background}
R~Cas is an O-rich Mira-type variable star with a period of 430.5 days. 
The star has a spectral type
of M7IIIe and a mean effective temperature 
$T_{\rm eff}\approx2500$~K (Haniff et al. 1995).
Using CO observations, Knapp et al. (1998) derived an expansion
velocity for the wind $V_{\rm out}=$12.1~\kms\ and a mass-loss
rate of 1.2$\times10^{-6}M_{\odot}$ yr$^{-1}$ (here we adopt a distance of 
160~pc derived from the period-luminosity relation by Haniff et
al. 1995).
Based on {\it IRAS} infrared observations, Young et al. (1993a,b)
found R~Cas is surrounded by an extended, dusty shell with an inner radius of
\am{1}{0} and outer radius \am{4}{3}. 
Recently, G\'erard \& Le~Bertre (2006) have reported a detection of
this star in \HI\ using the \nan\ telescope.

\subsubsection{Results: R Cas}
\begin{figure}
\epsscale{0.9}
\plotone{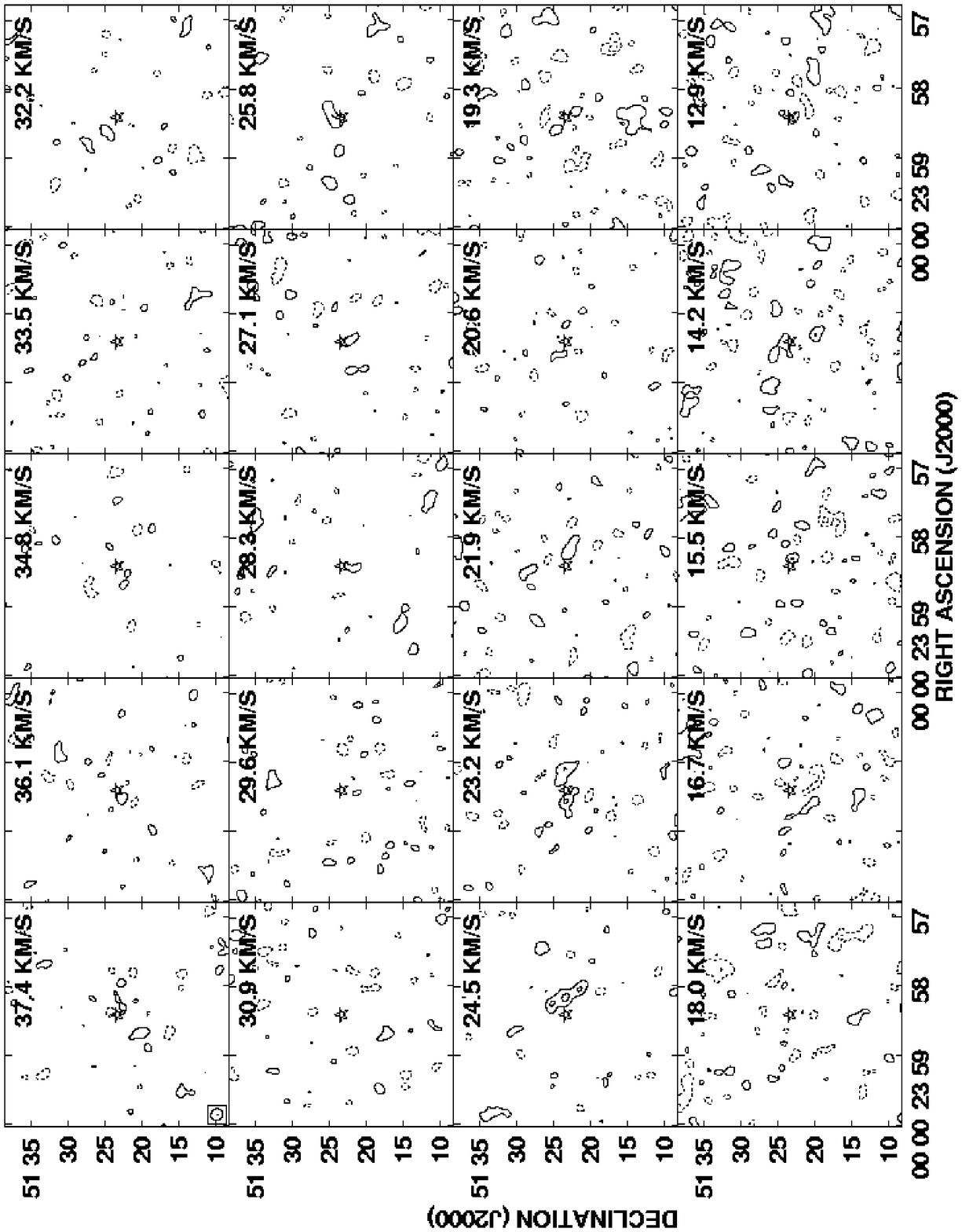}
\figcaption{\HI\ channel maps of the region around 
R~Cas. The maps have a spatial resolution of
$\sim101''\times92''$. Contour
levels are ($-$2,2,4)$\times$2.1~mJy beam$^{-1}$. The lowest
contour levels are $\sim2\sigma$. The systemic
velocity of the star is $V_{\rm sys,LSR}=24.9$~\kms, and a star symbol
marks its position. The range of channels shown corresponds to the
velocity range over which CO has been previously detected in the envelope of
R~Cas. 
\protect\label{fig:rcascmaps}}
\end{figure}

Figure~\ref{fig:rcascmaps} shows the \HI\ channel maps from our VLA
observations of R~Cas over a velocity range 
corresponding to the velocity spread of the CO(3-2) emission
detected in its envelope by Knapp et al. 
(1998) (12~\kms$\lsim V_{\rm LSR}\lsim$36~\kms). 
The systemic velocity of R~Cas derived from the CO observations is
$V_{\rm sys,LSR}=24.9\pm0.9$~\kms. 
Figure~\ref{fig:rcascmaps} shows that over this velocity interval,
the brightest detected \HI\ emission ($5\sigma$) is found in the channel
with central velocity 24.5~\kms---i.e., the channel closest
to the systemic velocity of the star.  
The emission peaks roughly one synthesized beam diameter 
from the optical
position of R~Cas (i.e., $\sim100''$ away). Two adjacent channels 
also show regions of extended emission within roughly 1-2 beam
diameters from the star. 

\begin{figure}
\plotone{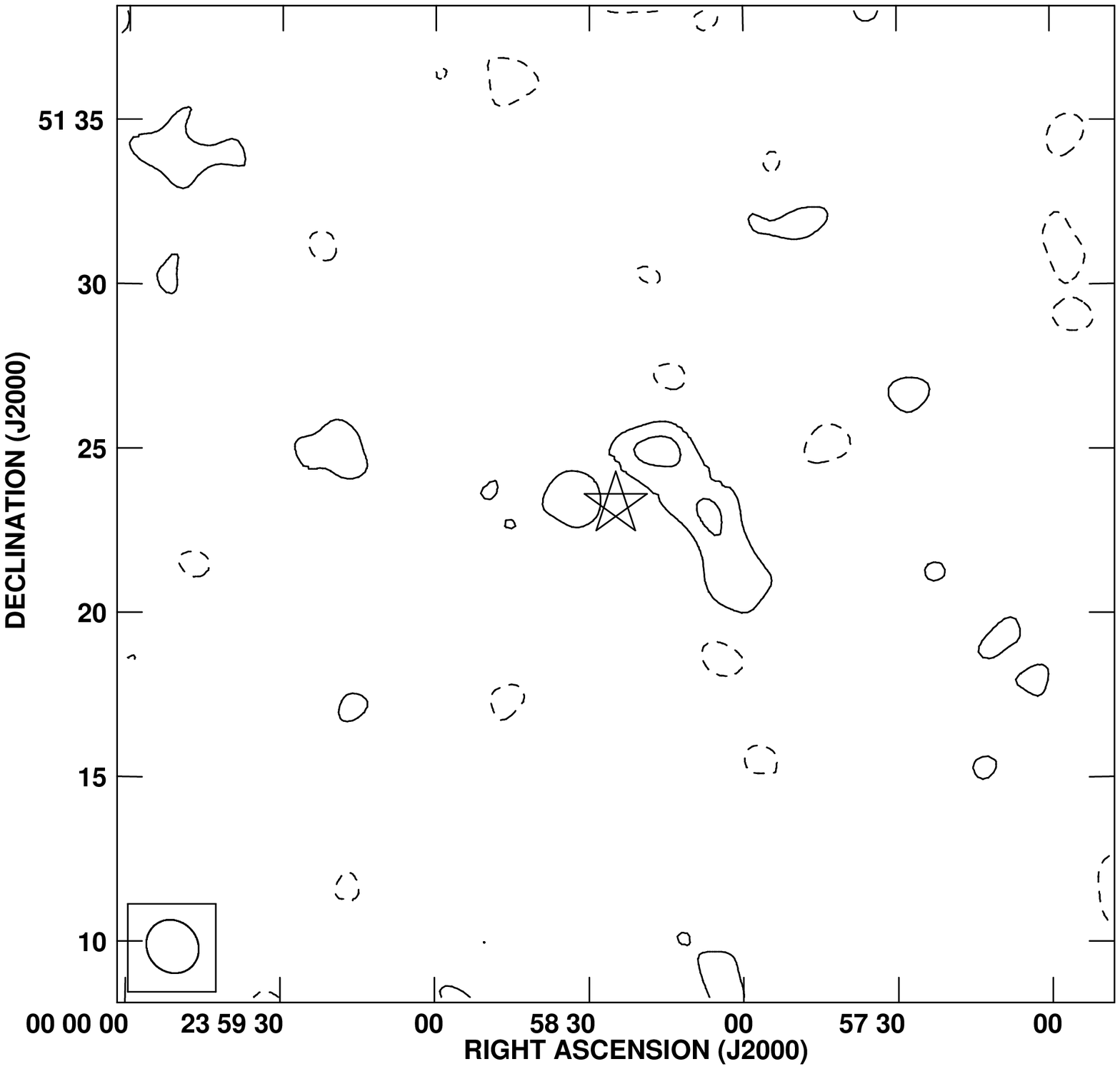}
\figcaption{\HI\ total intensity image of the region around
R~Cas, derived by summing the data over the velocity range 
$23.2$~\kms$\le V_{\rm sys,LSR}\le25.8$~\kms. 
Contour levels are ($-$2,2,4)$\times$5.4~Jy beam$^{-1}$
m s$^{-1}$. A star symbol marks the optical position of
R~Cas. The size of the region
shown is comparable to that of the FWHM of the primary beam. No 
correction for primary beam attenuation
has been applied.\protect\label{fig:rcasmom0}}
\end{figure}

\begin{figure}
\plotone{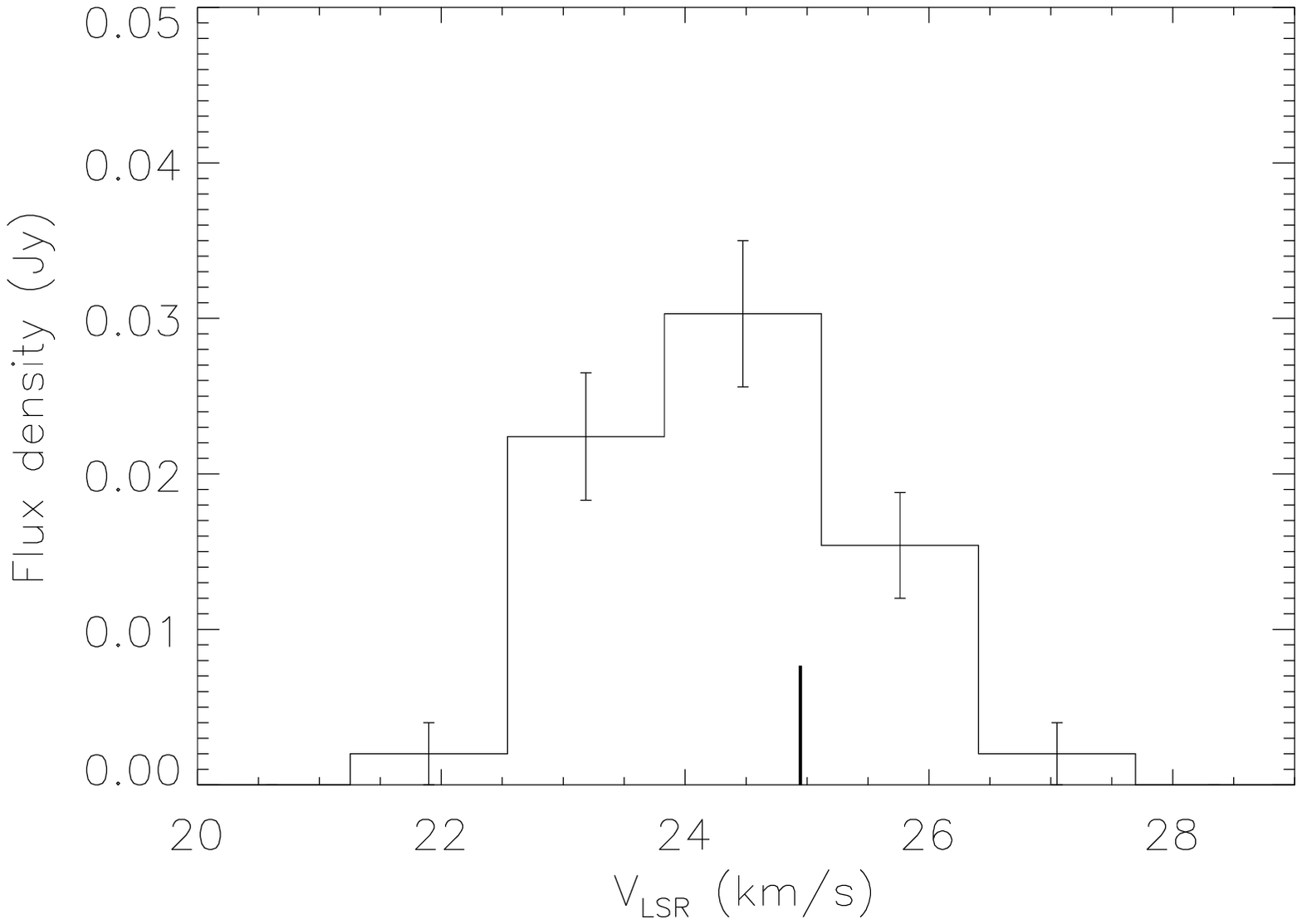}
\figcaption{Global (spatially integrated) 
\HI\ spectrum derived from emission detected toward
R~Cas. Error bars are as in Figure~\ref{fig:rscncglobal}.
The vertical bar marks 
the systemic velocity of the star derived from CO 
observations. CO has been detected toward R~Cas 
over the velocity interval $11$~\kms$\lsim V_{\rm LSR}
\lsim$38~\kms\ (Knapp et al. 1998).\protect\label{fig:rcasprofile}}
\end{figure}

Figure~\ref{fig:rcasmom0} shows an image formed from the sum
of three spectral channels spanning the velocity range 23.2~\kms$\le V_{\rm
LSR}\le$25.8~\kms. 
Because of the weakness of the emission and its narrow velocity
spread, no clipping or smoothing has been applied. The morphology
of the emission in Figure~\ref{fig:rcasmom0} suggests we may be seeing
a fragment of a clumpy, shell-like structure around R~Cas. If real,
this structure would overlap with the dust shell 
discovered by Young et al. (1993a,b) and would have a 
projected radius $r\gsim2.4\times10^{17}$~cm.
Using the blotch method described in \S~\ref{rscncresults}, 
we measure the total
\HI\ flux in this structure to be $S_{\rm HI}=0.087\pm0.007$~Jy~\kms.
At the distance of R~Cas, this corresponds to an 
\HI\ mass $M_{\rm HI}\approx5.3\times10^{-4}M_{\odot}$.  
A global \HI\ profile derived from these measurements is 
shown in Figure~\ref{fig:rcasprofile}. After correction for He,
the mass we estimate ($7.1\times10^{-4}M_{\odot}$) 
is still roughly an order of magnitude smaller than the
circumstellar envelope mass derived by Young et al. (1993b) from
infrared observations ($M=6.7\times10^{-3}M_{\odot}$). However, the dust
shell measured by Young et al. was approximately four times more
extended. Assuming a constant velocity wind, the density of
material, $\rho$, is expected to drop as the inverse square of the distance 
($\rho\propto r^{-2}$), implying that 
the total mass, $M$, will be proportional to $r$. 
Thus our estimate appears to be consistent with Young et
al.'s measurement if we are sampling only a small fraction of the envelope 
material.

\subsubsection{Discussion: An \HI\ Shell around R Cas?}
In our observations of R~Cas, roughly half of
our bandpass ($V_{\rm LSR}\lsim 10$~\kms, not shown in Figure~\ref{fig:rcascmaps}) 
is significantly contaminated by Galactic emission that is only
partially resolved out by the VLA. However, we find no evidence of
significant contamination in the higher-velocity
channels. Consistent with this, the spectra of Hartmann
\& Burton (1997) toward this direction show a steep drop-off in the
Galactic \HI\ brightness temperature 
near $V_{\rm LSR}\approx$15~\kms, and no detectable
Galactic emission at velocities $V_{\rm LSR}>$20~\kms\ (see also
G\'erard \& Le~Bertre 2006). 


We have  again used a matched filter search (see
\S~\ref{rscncresults}) as an aid in
quantifying the uniqueness and significance of the emission features detected in
our data cube. 
We searched  our tapered, naturally-weighted,
continuum-subtracted R~Cas data cube over a 30$'$
region. The search was limited to the velocity range $11.6\le V_{\rm
LSR}\le 59.2$~\kms\ in order to 
exclude edge channels and the portion of the band with
obvious Galactic contamination.
After smoothing the data in frequency using
Gaussian kernels with widths of 2-10 channels, we find the peak signal
over this search volume to
correspond spatially to the center of the
elongated structure in Figure~\ref{fig:rcasmom0}. The velocity
centroid of this feature occurs in the channel with center velocity
$V_{\rm LSR}=$23.2~\kms, 
and its peak signal-to-noise  (6$\sigma$) 
occurs for a smoothing kernel width of 4 channels
($\sim$5~\kms).  Our matched filter search turned up 
no other signals with significance $>5\sigma$ outside the channels
centered at $V_{\rm LSR}=$23.2~\kms and 24.5~\kms, respectively.
The results of this analysis are therefore 
consistent with the detection of \HI\ emssion from the circumstellar
envelope of R~Cas. 
We note also that the mean effective temperature of R~Cas
lies near the transition from
molecular to atomic winds proposed by Glassgold \& Huggins (1983),
implying that at least some atomic component to the wind is predicted for
this star.

The VLA \HI\ spectrum of R~Cas (Figure~\ref{fig:rcasprofile}) appears
quite similar to the single-dish profile recently published by 
G\'erard
\& Le~Bertre (2006) in terms of its central velocity, velocity width, and
peak flux density. However, it is unclear whether we have detected 
emission from the same material as those authors. 
G\'erard \& Le~Bertre find evidence that the \HI\
emission around R~Cas is quite extended (up to $\sim16'$) based on an
apparent increase in the measured flux density at the position of R~Cas
with increasing throw of the off-beams subtracted from the
on-source spectra. However, given
the weakness of the emission, estimates of the angular extent may be
influenced by baseline uncertainties in the single dish spectra.
Sensitive single-dish mapping of the region around 
R~Cas would likely provide additional insight.


\section{Summary and Concluding Remarks}
Recently, sensitive new single-dish \HI\ surveys have
established that neutral atomic hydrogen is common in the
circumstellar envelopes of evolved, low-to-intermediate mass 
stars undergoing mass-loss (G\'erard \& Le~Bertre 2006 and 
references therein). Studies of the
21-cm line emission from this material can therefore provide 
important constraints
on atmospheric models of AGB stars,
the physical conditions in their extended envelopes, and on
the  rates, timescales, and geometries of their mass-loss. 

Here we have reported the results of a VLA \HI\ imaging survey 
of five nearby AGB stars: RS~Cnc,
IRC+10216, EP~Aqr, R~Aqr, and R~Cas. 
\HI\ detections of four of these targets (RS~Cnc,
EP~Aqr, and R~Cas in emission and IRC+10216 in absorption) have been
published previously based on single-dish observations (Le~Bertre
\& G\'erard 2001,2004; G\'erard
\& Le~Bertre 2003,2006). However, because of limited spatial
resolution and confusion from Galactic emission along the
line-of-sight, the single-dish data alone did not
permit a full characterization of the small-scale structure of the
emission or its distribution relative to the star.

We have confirmed
the presence of \HI\ emission coincident in position and velocity with
the semi-regular variable
RS~Cnc, implying that the emission is indeed associated with its
circumstellar envelope. The emission comprises a compact,
slightly elongated region centered on the star with
a mean diameter  of $\sim82''$
($\sim1.5\times10^{17}$~cm),  plus an additional filament
extending $\sim6'$ to the northwest. 
We estimate a total \HI\ mass for this
material of
$M_{\rm HI}\approx 1.5\times10^{-3}M_{\odot}$.  
The morphology of this extended filament
suggests that a component of the mass-loss from RS~Cnc 
was highly asymmetric. From the \HI\ data
we derive a recent mass-loss rate of ${\dot
M}=1.7\times10^{-7}M_{\odot}$~yr$^{-1}$, comparable to
previous estimates based on CO observations.

For the Mira variable 
R~Cas we have detected weak emission centered at the systemic
velocity of the star. The  morphology of the emission is consistent with a partial
shell-like structure with a radius $r\sim100''$. This structure
overlaps with the dust shell previously detected by Young et
al. (1993a,b), and we estimate for it an \HI\ mass of $M_{\rm
HI}\approx 5.3\times10^{-4}M_{\odot}$. Galactic contamination at the
position and velocity of R~Cas is low, suggesting a good probability
that the \HI\ emission we have detected is associated with its
circumstellar envelope.

Toward two other targets (the carbon star IRC+10216 and the
semi-regular variable EP~Aqr) we have also detected multiple arcminute-scale
\HI\ emission features at velocities consistent with their respective circumstellar
envelopes, but spatially offset from the position of the stars. 
However, in these cases, 
we are unable to determine unambiguously whether the emission
arises from material within the circumstellar envelope or, instead, from the
chance superposition of
\HI\ clouds along the line-of-sight. In each case, 
we have discussed evidence for and
against both interpretations. 

Toward IRC+10216 we find arc-like \HI\ emission structures 
at projected distances of $r\sim14'-18'$ to the
northwest of the star. The large separation between the \HI\ emission and
the position of the star are consistent with the advanced
evolutionary status of IRC+10216 and the prediction that \HI\ will
be formed from an initially molecular wind via photodissociation
and/or the sweeping up of interstellar material as the wind expands. 
However, it is unclear if the highly asymmetric geometry and the complex 
velocity structure of the
emission we have detected are consistent with a circumstellar origin.

We were unable to
confirm the detection of \HI\ in absorption
against the cosmic background in the envelope of IRC+10216 
as previously reported by Le~Bertre \&
G\'erard (2001). Our VLA images reveal that 
at least part of the apparent absorption signature  seen
by Le~Bertre \& G\'erard may have arisen from emission that contaminated the
off-beam measurements used in their position-switched (on$-$off) observations.

In the case of EP~Aqr, we have detected six arcminute-scale
clumps of \HI\ emission within a projected radius of $\sim15'$ around
the star. All of
the clumps lie redward of the stellar systemic velocity, although
their velocities are consistent with the ``red
wing'' of the circumstellar envelope as seen in CO emission. While
it is tempting to posit that these clumps may be part of a more
diffuse underlying envelope reported by Le~Bertre \& G\'erard (2004), the
presence of an additional \HI\ clump in our data that is significantly
redshifted relative to the star raises the possibility that we are
instead sampling random interstellar clouds along the line-of-sight. We find,
however, that regardless of the assumed distance for these clumps,
their virial masses exceed their \HI\ masses by over an order of
magnitude, suggesting that either these are transient features or that
they are embedded in a more diffuse medium to which the VLA is insensitive.
In the case of both EP~Aqr and IRC+10216, combining our VLA
observations with single-dish mapping may help to remove the
ambiguities in associating the detected
\HI\ emission with the respective circumstellar envelopes. In
addition, a future compact ``E'' configuration of the VLA would be
well-suited to studies of this kind.

We detected our fifth target, R~Aqr (a symbiotic star with a hot
companion), in the 1.4~GHz continuum with a flux density
$F_{\rm 1.4GHz}=18.8\pm0.7$mJy. None of the other four stars in our
sample showed detectable continuum emission. R~Aqr is a well-known radio source,
and its continuum emission likely arises primarily from
free-free emission from an ionized circumbinary envelope.
However, we did not detect any neutral hydrogen associated with R~Aqr.

\begin{acknowledgements}
We thank E. G\'erard and T. Le~Bertre
for valuable discussions during the course of this work, and we thank
our referee, J. Knapp, for constructive comments that helped to improve
this paper.
LDM was supported by a Clay Fellowship and a Visiting Scientist
appointment from the Smithsonian
Astrophysical Observatory. This research has made use of the SIMBAD database,
operated at CDS, Strasbourg, France. The National Radio Astronomy
Observatory (NRAO) is a facility of the National Science Foundation 
operated under cooperative agreement by Associated Universities, Inc.

\end{acknowledgements}

\clearpage

\end{document}